\documentclass[showpacs, twocolumn,amsmath,amssymb, aps]{revtex4-1}

\usepackage[latin9]{inputenc}
\usepackage{verbatim}
\usepackage{amsmath}
\usepackage{amssymb}
\usepackage{graphicx}
\usepackage{wrapfig}
\usepackage{esint}
\usepackage{dcolumn}
\usepackage{bm}
\usepackage{setspace}
\usepackage{color}
\usepackage{soul}
\usepackage[T1]{fontenc}
\usepackage{lmodern}
\usepackage{subfig}

\begin{document}
\preprint{APS/123-QED}

\title{Magnetotransport of Weyl semimetals with tilted Dirac cones}
\author{Anirban Kundu}
\affiliation{Department of Electrical and Computer Engineering, National University of Singapore, 4 Engineering Drive 3, Singapore 117576}
\author{Hyunsoo Yang}
\affiliation{Department of Electrical and Computer Engineering, National University of Singapore, 4 Engineering Drive 3, Singapore 117576}
\author{M. B. A. Jalil}
\affiliation{Department of Electrical and Computer Engineering, National University of Singapore, 4 Engineering Drive 3, Singapore 117576}
\date{\today}


\begin{abstract}
Weyl semimetals (WSM) exhibit chiral anomaly in their magnetotransport due to broken conservation laws. Here, we analyze the magnetotransport of WSM in the presence of the time-reversal symmetry-breaking tilt parameter. The analytical expression for the magnetoconductivity is derived in the small tilt limit using the semiclassical Boltzmann equation. We predict a planar Hall current which flows transverse to the electric field and in the plane containing magnetic and electric fields and scales linearly with the tilt parameter. A tilt-induced transverse conductivity is also present in the case where the electric and magnetic fields are parallel to each other, a scenario where the conventional Hall current completely vanishes.

\end{abstract}

\maketitle

\section{Introduction}

A Weyl semimetal (WSM) state in topological systems with  broken
time-reversal and/or inversion symmetry can
be understood by the availability of the low energy quasiparticles called
Weyl fermions (WF) with linear in momentum dispersion near the crossing
point of two non-degenerate bands in momentum space known as Weyl
point (WP) \cite{1367-2630-9-9-356,PhysRevB.83.205101}. Although
the concept of WF was first introduced in relativistic field theory,
it has never been observed in the systems of elementary particles,
and has only very recently been realized in a condensed matter systems
as an excitation of quasi-particles. The essential properties of WPs
are: the presence of pairs of opposite chiralities connected by Fermi
arcs on the surface and the zero-sum of chiralities over all the WPs in
the Brillouin zone, according to the Nielsen-Ninomiya theorem
\cite{1983PhLB..130..389N}. The chirality $\chi$ of a WP  is
defined by the integration of Berry curvature $\mathbf{\Omega}_{\mathbf{k}}$
($\mathbf{\Omega}_{\mathbf{k}}=\nabla\times \mathbf{A}_{\mathbf{k}}$
with $\mathbf{A}_{\mathbf{k}}=i\left\langle u_{\mathbf{k}}|\nabla_{\mathbf{k}}u_{\mathbf{k}}\right\rangle $,
where $\left|u_{\mathbf{k}}\right\rangle $ is the Bloch wave
function) over a closed surface around that WP in momentum space ($\varointclockwise\text{d}\mathbf{s}\cdot\mathbf{\Omega}_{\mathbf{k}}=2\pi\chi$
where, $\chi=\pm1,\pm2,.....$ ).

In the presence of  external electric and magnetic fields the conservation of the number of WFs with particular chirality is broken, a phenomenon known as chiral anomaly. The WSMs are distinctive from other metals/semi-metals because the
chiral anomaly can lead to unusual magneto-transport phenomena.
For example; the observation of negative magneto-conductivity has been attributed to the phenomenon of chiral anomaly  \cite{Arnold2016,Li2015,Li2016,PhysRevX.5.031023,Xiong413,Zhang2016}.

\begin{figure}
\begin{centering}
\subfloat[]{\includegraphics[width=0.5\linewidth]{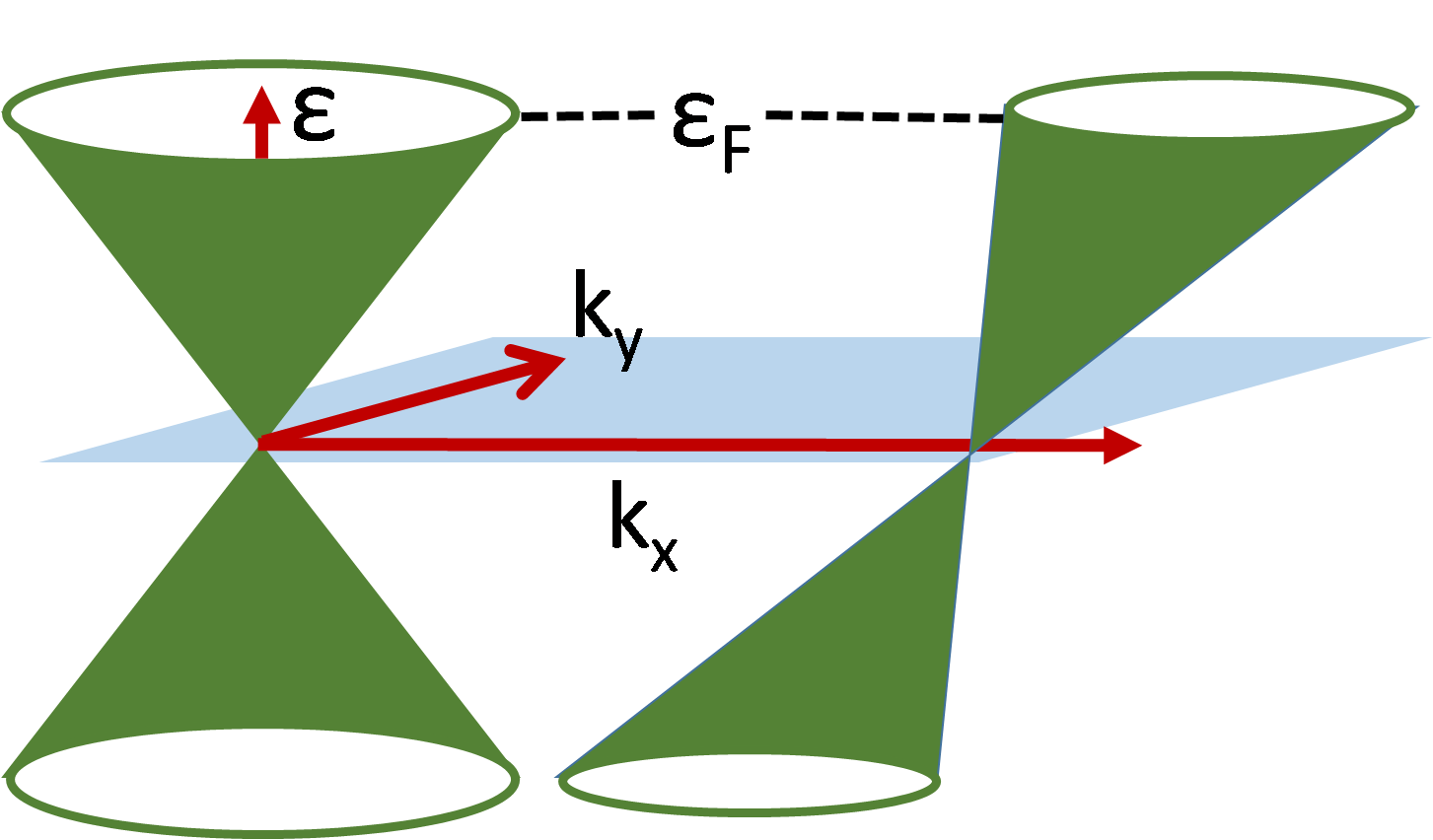}}
~~~~~~~~~~\subfloat[]{\includegraphics[width=0.35\linewidth]{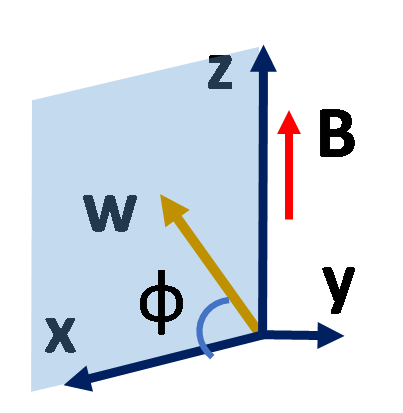}}
\vfill{}
\subfloat[]{\includegraphics[width=0.5\linewidth]{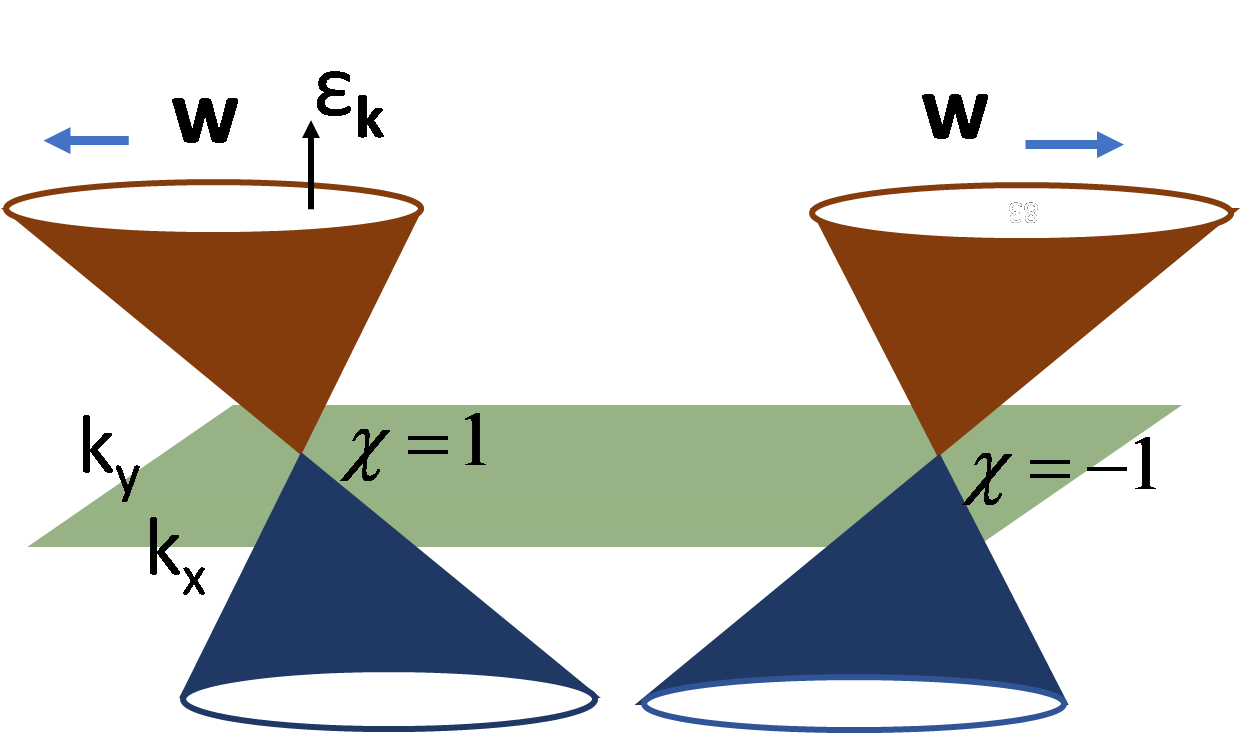}}
\subfloat[]{\includegraphics[width=0.5\linewidth]{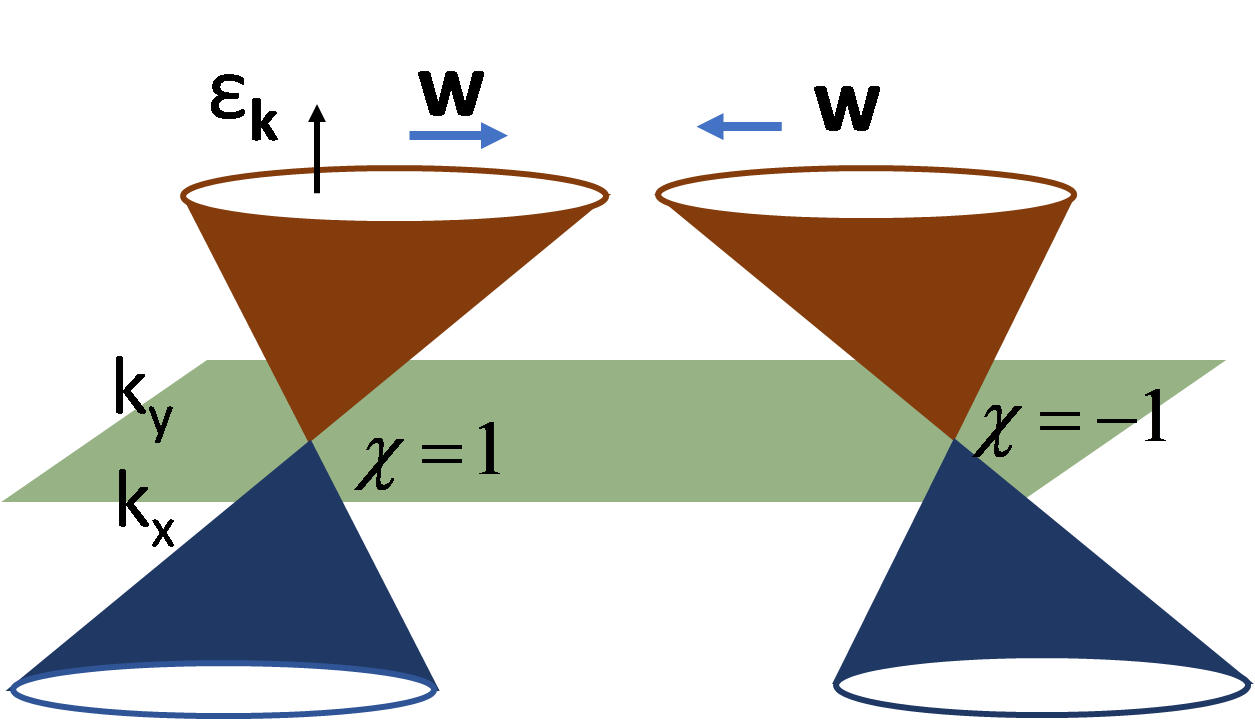}}
\par\end{centering}
\caption{\label{direction1} (Color online) 
(a) Dispersion relation for the non-tilted (left) and tilted (right) WSM.
(b) The $z$-axis is parallel to the magnetic field while the tilt vector $\mathbf{w}$ lies on the $x$-$z$ plane with an angle $\phi$ from the $x$-axis. 
(c)-(d), A pair of  Dirac cones are tilted (c) towards each other, and
(d)  away from each other.
}
\end{figure}

In this manuscript, we study the electron transport in WSM where the Dirac cones are tilted in momentum space. Such a tilted WSM can be described by an additional anisotropy term in
the Hamiltonian for electrons near the WPs as follows \cite{PhysRevB.78.045415},
\begin{equation}
H=v_F\boldsymbol{\sigma}\cdot\mathbf{k}+I\mathbf{w}\cdot\mathbf{k},\label{eq: model hamiltonian}
\end{equation}
where the components of $\boldsymbol{\sigma}$ are  the Pauli matrices and $I$
is the $2\times2$ identity matrix, $\mathbf{k}$ is the wave vector, the vector $\mathbf{w}$ represents the dispersion tilt
and $v_F$ is the Fermi velocity. The tilt $\mathbf{w}$ in Eq. (1) breaks the time-reversal symmetry
(TRS).  The energy eigenvalues are given by
$\epsilon_{k}=(\hbar\mathbf{w}\cdot\mathbf{k}+\hbar v_F \mathrm{k}$) and the corresponding group velocity is
$\mathbf{v_{k}}=(\mathbf{w}+v_F\hat{\mathrm{k}})$ (where $\hat{\mathrm{k}}$
is the unit vector along $\mathbf{k}$ and  $v_F$  the Fermi velocity). 
If the value of tilt
$|\mathbf{w}|<v_F$  then
the WSMs are sub-grouped as Type-I WSM and when $|\mathbf{w}|>v_F$
then the WSMs are sub-grouped as Type-II WSM.  
A non-tilted and tilted  dispersion relation in WSM is shown in the schematic diagram in Fig. \ref{direction1}(a).
Type-II
WSM phase has been observed in a range of materials such as MoTe$_{2}$,
WTe$_{2}$, LaAlGe \cite{Soluyanov2015,Xue1603266,Deng2016,Li2017}
and already several transport phenomena have been addressed in several  studies \cite{PhysRevB.95.075133,PhysRevLett.117.077202,APhysRevB.94.081408,Zyuzin2016}. 

Transport
studies of tilted Weyl semimetals   have been done recently  such
as anomalous Nernst effect \cite{PhysRevB.96.115202,Saha2018},
optical response to circularly polarized light 
 \cite{PhysRevB.96.085114} and planar Halle effect (PHE) \cite{PhysRevB.99.115121}. Using Boltzmann transport
it has been predicted that both the chiral anomaly and non-trivial
Berry curvature effects leads to the PHE in Weyl semimetals \cite{PhysRevLett.119.176804} and very recently,
 PHE  in half Heusler
Weyl semimetal GdPtBi has been attributed to a strong Berry curvature effect \cite{PhysRevB.98.041103}. 

Motivated by the above 
works, we  analyze the role of the dispersion
tilt in Berry curvature induced transport in the WSM systems,
which has not been investigated hitherto. 
Our study reveals  unusual transport properties, such as the planar Hall effect induced by the time-reversal symmetry breaking tilt and Berry curvature effects.
 We mainly consider tilted Type-I WSM and evaluate transport properties in 
 response to external electric and magnetic fields.
 Both the longitudinal and transverse conductivities are calculated
using the Boltzmann transport equation. Our manuscript
is arranged as follows: 
in Sec. II we present the general framework of the semiclassical Boltzmann model to calculate the current density up to 
the second orders in the $B$ field. 
  in Sec. III we present the analytical expression for the conductivity in the small-tilt regime. We also perform numerical calculations for arbitrary tilt value in the case of both Type-I and Type-II WSM (results of the latter are presented in more detail in the Supplemental) and verify the analytical results in the limit of small tilt. The anisotropy of the WSM conductivity with respect to the magnetic field and tilt directions is also presented.
  We conclude in Sec. IV.

\section{Method}

We apply the semiclassical Boltzmann transport model under the assumption of  a small external magnetic and electric fields where the separation between Landau
levels can be neglected and where  it is valid to
use the semiclassical approach \cite{PhysRevB.88.104412}. We begin with
the standard procedure to calculate the Boltzmann's distribution function
for a system with non-zero Berry curvature under the application of
external electric and magnetic field. The presence of Berry curvature
$\mathbf{\Omega_{k}}$ in the WSMs provides a correction to the
phase space volume in the case of adiabatic transport, as shown in
\cite{PhysRevB.59.14915,RevModPhys.82.1959}. In the presence of electric
($\mathbf{E}$) and magnetic field ($\mathbf{B}$) the semiclassical
equations of motion are modified as follows \cite{PhysRevB.59.14915},
\begin{align}
\dot{\mathbf{r}} & =\left(1+\frac{e}{\hbar}\mathbf{\Omega}_{\mathbf{k}}\cdot\mathbf{B}\right)^{-1}\left[\mathbf{v_{k}}+e\mathbf{E}\times\mathbf{\Omega}_{\mathbf{k}}+\frac{e}{\hbar}(\mathbf{\Omega}_{\mathbf{k}}\cdot\mathbf{v}_{\mathbf{k}})\mathbf{B}\right],\nonumber \\
\hbar\dot{\mathbf{k}} & =\left(1+\frac{e}{\hbar}\mathbf{\Omega}_{\mathbf{k}}\cdot\mathbf{B}\right)^{-1}\left[e\mathbf{E}+\frac{e}{\hbar}\mathbf{v_{k}}\times\mathbf{B}+\frac{e^{2}}{\hbar}(\mathbf{E}\cdot\mathbf{B})\mathbf{\Omega}_{\mathbf{k}}\right],\label{eq:smcl}
\end{align}
where $\mathbf{v_{k}}=\partial\epsilon_{\mathbf{k}}/\hbar\partial\mathbf{k}$
is the group velocity of the electrons with $\epsilon_{\mathbf{k}}$
being the energy dispersion, $\mathbf{r}$ the position and $\hbar\mathbf{k}$
the momentum of a single electron.
 To evaluate the distribution $f(\mathbf{r},\mathbf{k},t)$
we consider the Boltzmann's equation, 
\begin{equation}
\frac{\partial f(\mathbf{r},\mathbf{k},t)}{\partial t}+\dot{\mathbf{r}}\cdot\frac{\partial f(\mathbf{r},\mathbf{k},t)}{\partial\mathbf{r}}+\dot{\mathbf{k}}\cdot\frac{\partial f(\mathbf{r},\mathbf{k},t)}{\partial\mathbf{k}}=I_{coll}\left\{ f(\mathbf{r},\mathbf{k},t)\right\} ,\label{eq:boltzman}
\end{equation}
where $I_{coll}\left\{ f(\mathbf{r},\mathbf{k},t)\right\} $
is the collision integral. 
We solve for the change in the distribution function $\delta f$ for a uniform system under steady-state condition, i.e., ignoring the time and space dependence in the above equation, and apply the relaxation time approximation, i.e., $I_{coll}\left\{ f(\mathbf{r},\mathbf{k},t)\right\} =\delta f(\mathbf{r},\mathbf{k},t)/\tau(\mathbf{k})$.
Here, $\tau(\mathbf{k})$
is the relaxation time and is assumed to be independent of $\mathbf{k}$. This is a common assumption given that more refined forms of $\tau(\mathbf{k})$ do not lead
to significantly different physics \cite{PhysRevB.88.104412}. Assuming the second-order derivative
of the distribution function to be negligible, the change in the 
distribution function is evaluated from Eqs. (\ref{eq:smcl}) and (\ref{eq:boltzman})
as, 
\begin{align}
\delta f(\mathbf{r},\mathbf{k},t) & =-\tau\left(1+\frac{e}{\hbar}\mathbf{\Omega}_{\mathbf{k}}\cdot\mathbf{B}\right)^{-1}\hbar^{-1}\nonumber \\
 & \left(e\mathbf{E}+\frac{e^2}{\hbar}\left(\mathbf{E}\cdot\mathbf{B}\right)\mathbf{\Omega}_{\mathbf{k}}\right)\cdot\mathbf{v_{k}}\left(\partial f^{0}/\partial\epsilon\right),
\end{align}
where $\epsilon$ is the electron energy and $f^{0}$ is the equilibrium distribution function which usually
can be replaced by Fermi function. As we are interested in the Berry-curvature
induced effect only,  we neglect the effect of Lorentz force (see the second line of Eq. (\ref{eq:smcl}))
which produces the conventional Hall current. The expression of current
density is given by, $\mathbf{j}=\left(e/8\pi^{3}\right)\int\text{d}\mathbf{k}\text{ }(1+(e/\hbar)\mathbf{\Omega}_{\mathbf{k}}\cdot\mathbf{B})^{-1}\dot{\mathbf{r}}\delta f(\mathbf{k})$
and using the above results for a single Weyl node we obtain,

\begin{align}
\mathbf{j} & =\frac{e\tau}{4\pi^{3}\hbar}\int\text{d\ensuremath{\epsilon}}\frac{\partial f^{0}}{\partial\epsilon}\int\text{d}S\frac{1}{(1+\frac{e}{\hbar}\mathbf{\Omega}_{\mathbf{k}}\cdot\mathbf{B})|\mathbf{v_{k}}|}\left[(e\mathbf{E}\cdot\mathbf{v_{k}})\mathbf{v_{k}}\right.\nonumber \\
 & +\frac{e^{2}}{\hbar}(\mathbf{E}\cdot\mathbf{B})\left((\mathbf{\Omega}_{\mathbf{k}}\cdot\mathbf{v_{k}})\mathbf{v_{k}}+\frac{e\mathbf{B}}{\hbar^{2}}(\mathbf{\Omega}_{\mathbf{k}}\cdot\mathbf{v_{k}})^2\right)\nonumber \\
 & \left.+\frac{e^{2}\mathbf{B}}{\hbar}(\mathbf{E}\cdot\mathbf{v_{k}})(\mathbf{\Omega}_{\mathbf{k}}\cdot\mathbf{v_{k}})\right],
\end{align}
 where $S$  is the  area of the constant energy surface. The current density above originates due to the imbalance between the particles numbers in the right-handed and left-handed valleys in the presence of the external electric and magnetic fields (i.e., the phenomenon  of  chiral anomaly), as discussed in Ref. \cite{PhysRevB.88.104412}. Next, we expand $\mathbf{j}$ to various orders of magnetic field
as follows: $\left(1+(e/\hbar)\mathbf{\Omega}_{\mathbf{k}}\cdot\mathbf{B}\right)^{-1}=1-(e/\hbar)\mathbf{\Omega}_{\mathbf{k}}\cdot\mathbf{B}+(e/\hbar)^2\left(\mathbf{\Omega}_{\mathbf{k}}\cdot\mathbf{B}\right)^{2}+\mathcal{O}\left(|\mathbf{B}|^{3}\right)$
and so that one can express  the current density as, $\mathbf{j}=\mathbf{j}^{(0)}+\mathbf{j}^{(1)}+\mathbf{j}^{(2)}+...$
where the numbers in the superscript represent the orders of the magnetic
field. Next, we substitute the expression of the Berry curvature as
given by, $\mathbf{\Omega}_{\mathbf{k}}=\left(\mathrm{k}_{F}^{2}/2\mathrm{k}^{3}\right)\mathbf{k}$
(where $k_{F}$ is related to the Fermi energy by $\epsilon_{F}=\hbar v_F \mathrm{k}_{F}$)
\cite{PhysRevB.95.075133} into the above expression
of current density and obtain for a single valley,

\begin{equation}
\mathbf{j}^{(0)}=\sigma_{0}\int\text{d\ensuremath{\epsilon}}\frac{\partial f^{0}}{\partial\epsilon}\int\frac{\text{d}S}{k_{F}^{2}|\mathbf{v_{k}}|v_F}\left(\mathbf{E}\cdot\mathbf{v_{k}}\right)\mathbf{v_{k}},\label{eq:current_zeroth_B}
\end{equation}

\begin{align}
\mathbf{j}^{(1)} & =\sigma_{0}c_{b}\int\text{d\ensuremath{\epsilon}}\frac{\partial f^{0}}{\partial\epsilon}\int\frac{\text{d}S}{k_{F}^{2}|\mathbf{v_{k}}|v_F}\left[\mathbf{B}\left(\mathbf{E}\cdot\mathbf{v_{k}}\right)\left(\mathbf{\Omega}_{\mathbf{k}}\cdot\mathbf{v_{k}}\right)\right.\nonumber \\
 & \left.+\left(\mathbf{E}\cdot\mathbf{B}\right)\left(\mathbf{\Omega}_{\mathbf{k}}\cdot\mathbf{v_{k}}\right)\mathbf{v_{k}}-\left(\mathbf{\Omega}_{\mathbf{k}}\cdot\mathbf{B}\right)\left(\mathbf{E}\cdot\mathbf{v_{k}}\right)\mathbf{v_{k}}\right],\label{eq:current_first_B}
\end{align}

\begin{align}
\mathbf{j}^{(2)} & =\sigma_{0}c_{b}^{2}\int\text{d\ensuremath{\epsilon}}\frac{\partial f^{0}}{\partial\epsilon}\int\frac{\text{d}S}{k_{F}^{2}|\mathbf{v_{k}}|v_F}\left[\left(\mathbf{\Omega}_{\mathbf{k}}\cdot\mathbf{B}\right)^{2}\left(\mathbf{E}\cdot\mathbf{v_{k}}\right)\mathbf{v_{k}}\right.\label{eq:current_second_B}\nonumber \\\
 & -\mathbf{B}\left(\mathbf{E}\cdot\mathbf{v_{k}}\right)\left(\mathbf{\Omega}_{\mathbf{k}}\cdot\mathbf{B}\right)\left(\mathbf{\Omega}_{\mathbf{k}}\cdot\mathbf{v_{k}}\right)\nonumber \\
 & \left.+\left(\mathbf{E}\cdot\mathbf{B}\right)\left(\mathbf{B}\left(\mathbf{\Omega}_{\mathbf{k}}\cdot\mathbf{v_{k}}\right)^{2}\frac{\mathbf{v_{k}}}{|\mathbf{v_{k}}|}-\left(\mathbf{\Omega}_{\mathbf{k}}\cdot\mathbf{v_{k}}\right)\left(\mathbf{\Omega}_{\mathbf{k}}\cdot\mathbf{B}\right)\mathbf{v_{k}}\right)\right].\nonumber \\
\end{align}
In the above,  $c_{b}=e\hbar v_F^{2}/\epsilon_{F}^{2}$ and
$\sigma_{0}=e^{2}\tau\epsilon_{F}^{2}/4\pi^{3}\hbar^{3}v_F$. Note that
the first and second-order terms in $B$  arise entirely as a result of the non-zero Berry curvature. Now, to evaluate the current density
one needs to integrate over energy and momentum. We consider a model
Hamiltonian given in Eq. (\ref{eq: model hamiltonian}) with linear
dispersion which describes a tilted WSM (for both Type-I and Type-II). For the Type-I case, we calculate the current density in two
different ways: i) we first assume a small tilt limit and obtain the
analytical expression of the current density to the first order in
the tilt vector $\mathbf{w}$; ii), we calculate
the current density numerically  for arbitrary value of tilt, and compare the
results with the analytical expression obtained in the small tilt
limit. For Type-II WSM,  we only perform the numerical calculation
and present the results in the  Supplemental.

\section{Tilted WSM}

\begin{figure}
\begin{centering}
\subfloat[]{\includegraphics[width=0.5\linewidth]{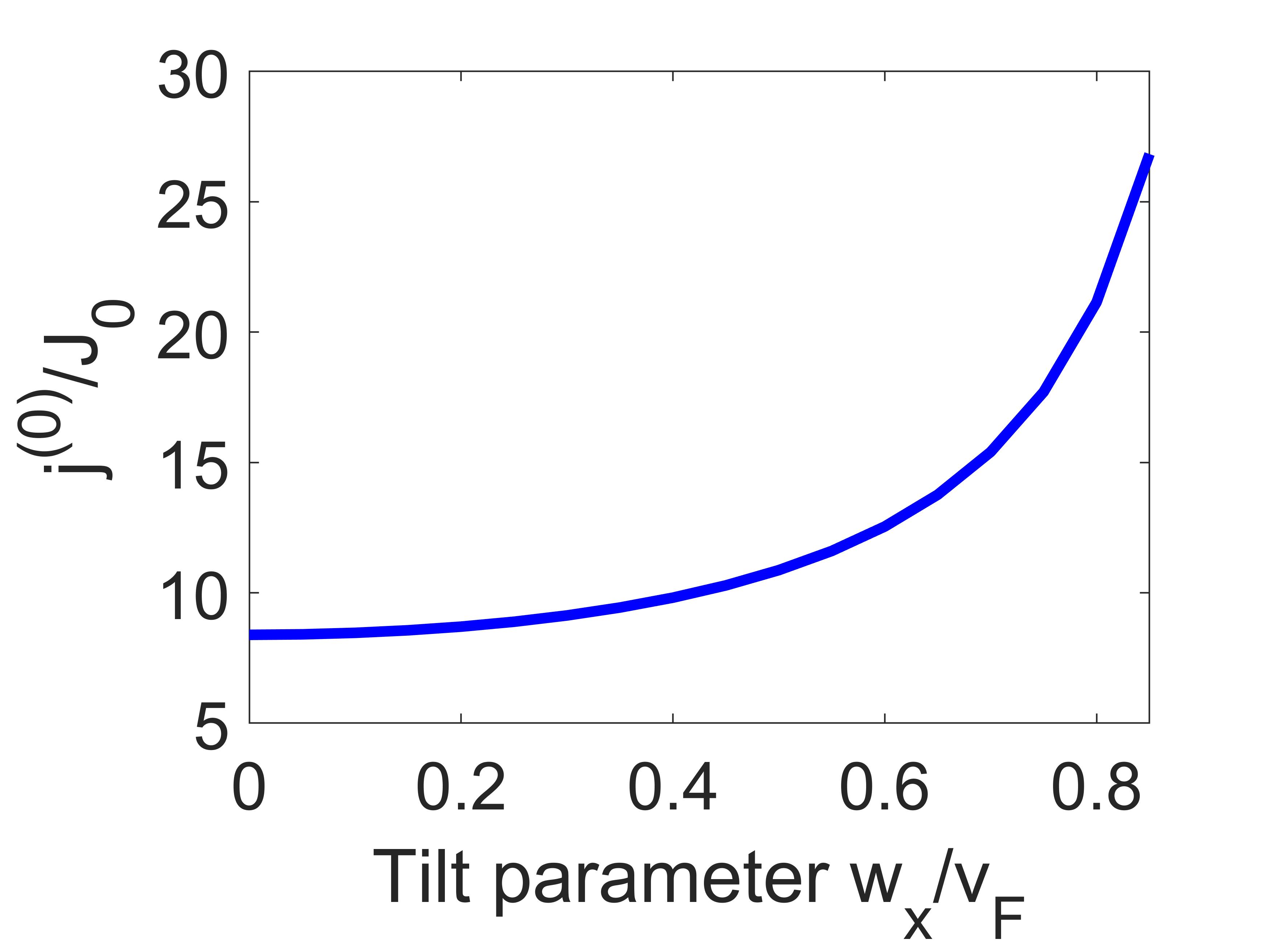}}
\subfloat[]{\includegraphics[width=0.5\linewidth]{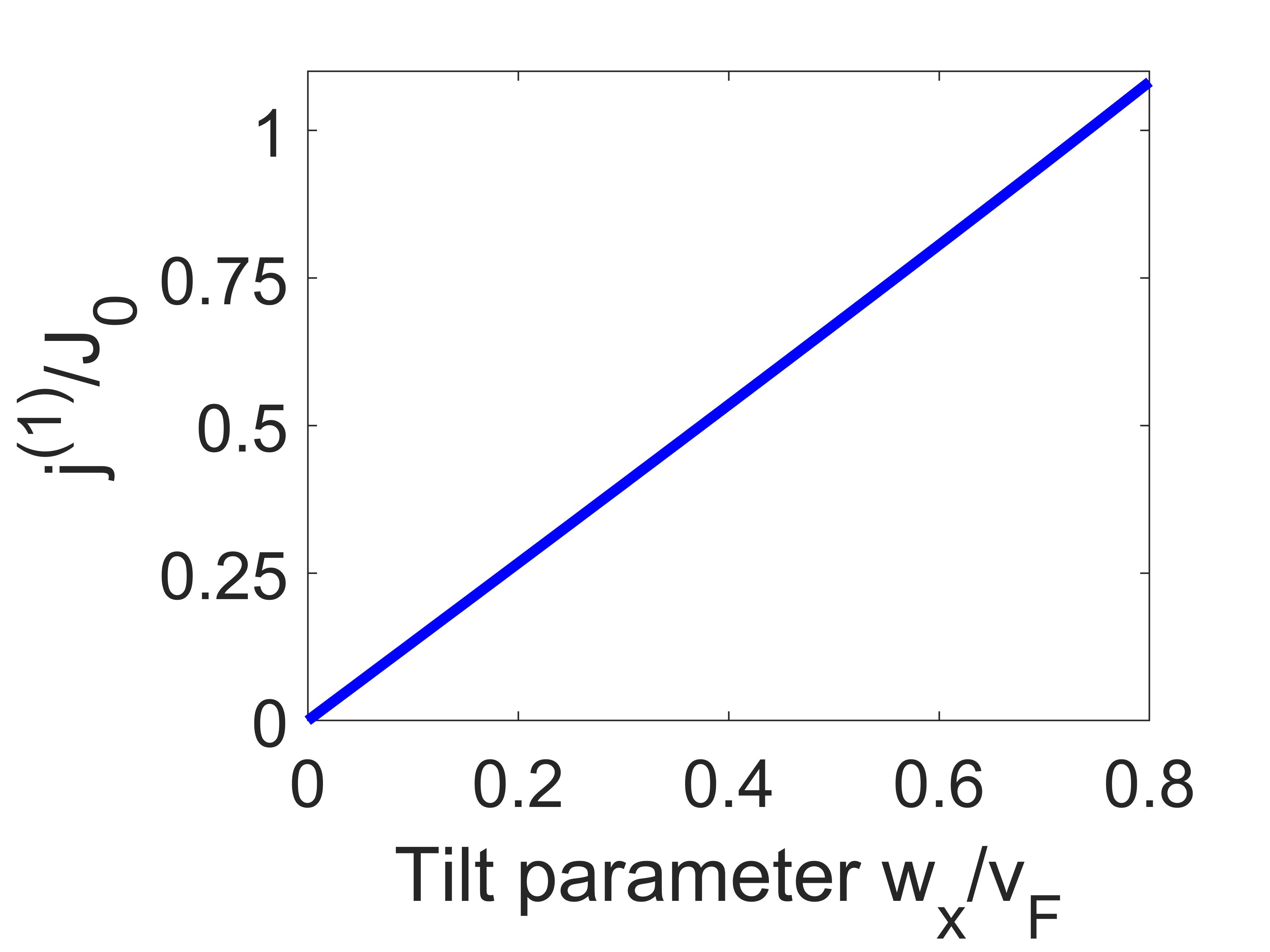}}
\vfill{}
\subfloat[]{\includegraphics[width=0.5\linewidth]{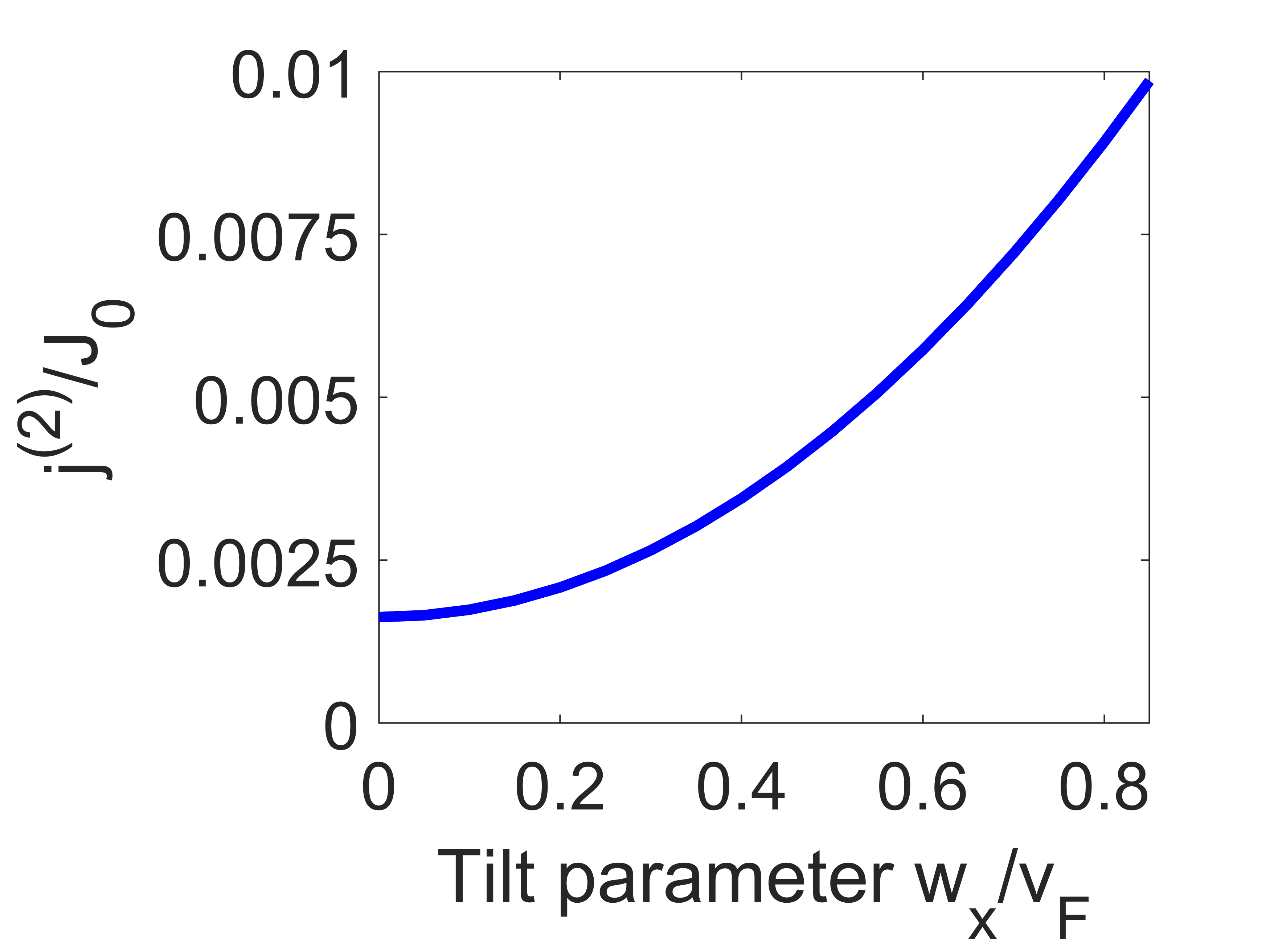}}
\subfloat[]{\includegraphics[width=0.5\linewidth]{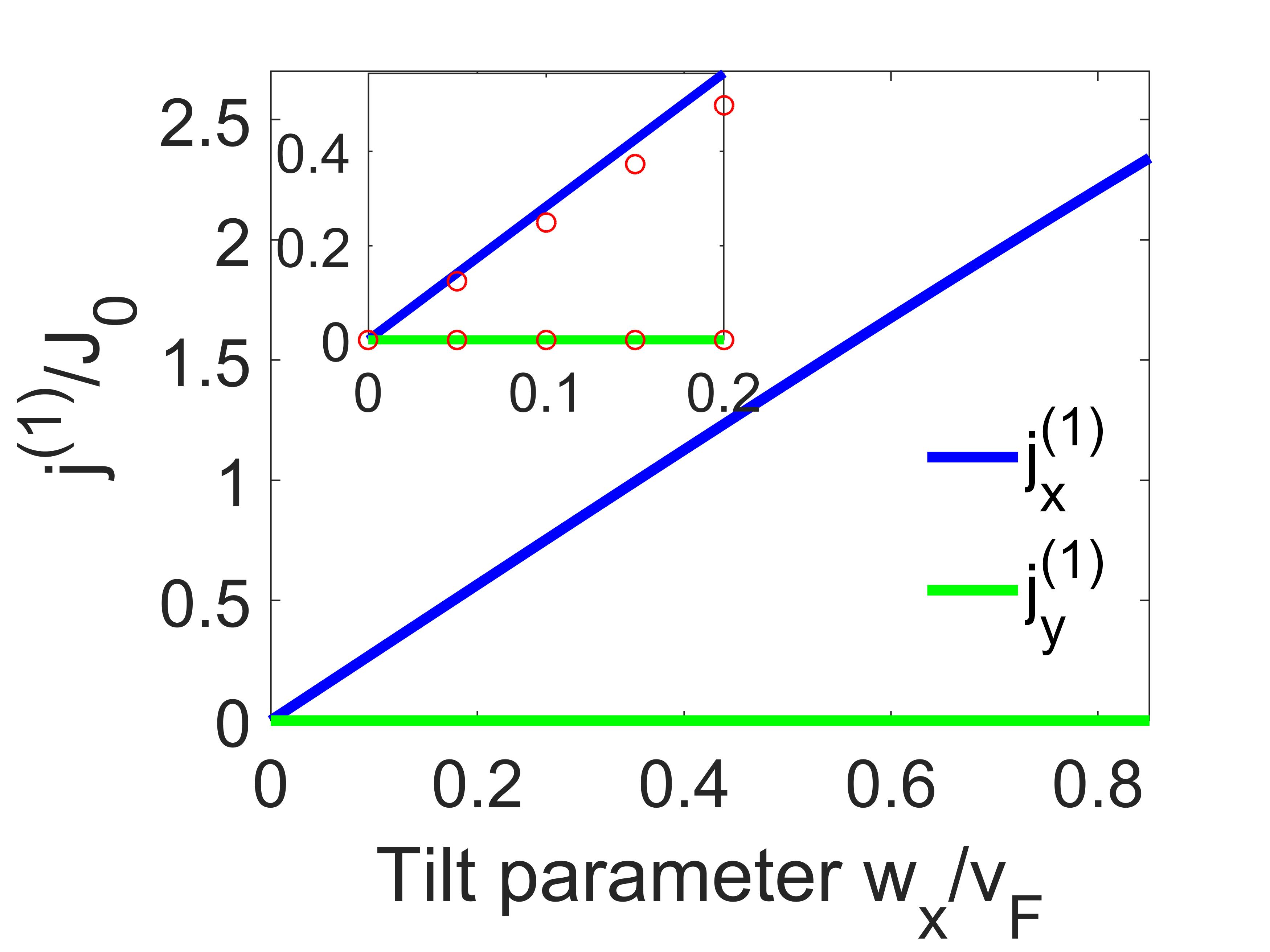}}

\par\end{centering}
\caption{\label{Fig:j1_hall_signal} (Color online)
 (a)-(c), Plot of longitudinal components of current density $\mathbf{j}^{(0)}$,
$\mathbf{j}^{(1)}$ and $\mathbf{j}^{(2)}$ with respect to
$x$-component of tilt ($\text{w}_{x}$) when electric field $\mathbf{E}\parallel (\hat{x}+\hat{z})$. We use $J_0=\sigma_0 |\mathbf{E}|$.
(d) Plot of $x$ and $y$
components of $\mathbf{j}^{(1)}$ with respect to $x$-component
of tilt $\text{w}_{x}$ when $\mathbf{E}||\mathbf{B}||\hat{z}$.  Inset: Comparison
with the analytical results (red circle), which are in close agreement with the numerical results at the small tilt limit.
}
\end{figure}
\textit{ Small tilt limit} ($|\mathbf{w}|\ll v_F$): 
For small tilt, we expand Eqs. (\ref{eq:current_zeroth_B}),
(\ref{eq:current_first_B}) and (\ref{eq:current_second_B}) to the
first order in $\mathbf{w}$ using $|\mathbf{v_{k}}|\approx v_F+(\hat{\mathrm{k}}\cdot\mathbf{w})+\mathcal{O}\left(|\mathbf{w}|^{2}\right)$.
The presence of a finite tilt to the energy dispersion changes the shape
of the Fermi surface (FS). However, in the small tilt limit, one can neglect
this change and assume the shape of the FS to be spherical. We also consider
the low-temperature limit such that we can approximate the distribution
function $f^{0}$ to be the step function $\theta(\epsilon-\epsilon_{F})$
and consequently $\partial f^{0}/\partial\epsilon$ can be replaced
by a Dirac delta function given by $\delta(\epsilon-\epsilon_{F})$
(note that most of the experimental  transport measurements on WSMs are  done at low temperatures at which this approximation is valid). Without loss
of any generality, we choose the $z$-axis to be aligned along the
magnetic field ($\mathbf{B}=B\hat{z}$) and that $\mathbf{w}$
and $\mathbf{B}$ lie in the same plane (Fig. \ref{direction1}(b)). Choosing the plane to
be the $x$-$z$ plane, we write, $\mathbf{w}=\text{w}_\text{{x}}\hat{x}+\text{w}_{z}\hat{z}$
($\text{w}_{x}$ and $\text{w}_{z}$ are the components of tilt and $\hat{x}$ and
$\hat{z}$ are unit vectors along $x$ and $z$-axis ). 
Under the
above assumptions, we calculate the current densities for a pair of
Weyl nodes with opposite chirality to the first order in tilt and
to the various orders in magnitude of the magnetic field $B$. While doing
so we considered two distinct cases: (I) the tilt changes sign between
the valleys with opposite chirality i.e. $\mathbf{w}_{\chi}=\chi\mathbf{w}$,
and (II) the sign of the tilt remains unchanged in the two valleys i.e.
$\mathbf{w}_{\chi}=\mathbf{w}$.

\textit{Case (a):} In the small tilt limit,  the current density can be evaluated  analytically derived from the integrals in Eqs. \eqref{eq:current_zeroth_B}-\eqref{eq:current_second_B}  with $\mathbf{E}$ is chosen to be along $\parallel (\hat{x}+\hat{z})$.  
The current density of different orders in $B$ is given by (see the Supplemental Information for more details on the derivation): 
\begin{align}
\mathbf{j}^{(0)} & =\frac{8\pi}{3}\sigma_{0}\mathbf{E},\label{eq:chiraltilt_zeroth}
\end{align}
\begin{align}
\mathbf{j}^{(1)} & =\frac{4\pi}{15}\sigma_{0}c_{b}\left[18\left(\mathbf{E}\cdot\mathbf{B}\right)\mathbf{w}+19\left(\mathbf{E}\cdot\mathbf{w}\right)\mathbf{B}\right.\nonumber \\
 & \left.+\hat{x}\left(\mathbf{E}\times\mathbf{w}\cdot\hat{y}\right)B-\hat{y}\left(\mathbf{E}\times\mathbf{w}\cdot\hat{x}\right)B\right],\label{eq:chiraltilt_first}
\end{align}

\begin{align}
\mathbf{j}^{(2)} & =2\pi\sigma_{0}c_{b}^{2}\left(\frac{1}{3}\left(\mathbf{E}\cdot\mathbf{B}\right)\mathbf{B}+\frac{1}{15}\mathbf{E}\right).\label{eq:chiraltilt_second}
\end{align}
Interestingly, the first-order term is non-vanishing only in the presence of
tilt, while the zeroth and second-order terms are independent of the
tilt. This tilt dependence can be understood
from the following argument. The anisotropic tilt appears in the integrals  in Eqs. (\ref{eq:current_zeroth_B}) to (\ref{eq:current_second_B}) because of the dependence of the group velocity $\mathbf{v_{k}}$ on tilt. In the case of both $\mathbf{j}^{(0)}$ and
$\mathbf{j}^{(2)}$ all the terms in the integral are even
in $\mathbf{k}$ in absence of tilt. However, in the presence of the dispersion tilt,
these terms (to the first order in tilt) become odd in $\mathbf{k}$ i.e. the integrals change sign as one goes from $\mathbf{k}$ to $-\mathbf{k}$,
and, as a consequence, vanish after integration over the FS
(which is taken to be  spherical). Conversely, for $\mathbf{j}^{(1)}$,
all the terms in the current integral up to the first order in tilt are even
in $\mathbf{k}$, and hence integration over the FS is not necessarily
zero.\textcolor{red}{{}}

To understand the dependence of tilt on the current, we consider the implications of   Eq. (\ref{eq:chiraltilt_first}).
We consider two particular cases, first, $\mathbf{E}\parallel\mathbf{B}$
i.e. the measurement set-up for longitudinal magnetoconductivity  and second, $\mathbf{E}\perp\mathbf{B}$
i.e. the usual Hall measurement set-up. In the first case, i.e. for $\mathbf{E}\parallel\mathbf{B}\parallel\hat{z}$,
we obtain, 
\begin{align}
\mathbf{j}^{(1)} & =\sigma_{0}c_{b}\frac{4\pi}{15v_F}\left(19B\mathbf{w}+18\mathrm{w}_{z}\mathbf{B}\right)E.\label{eq:firstorderchiral3}
\end{align}
Clearly, the current density has components not only along the direction of the magnetic field but also along the tilt directions. As a consequence, we have current transverse to the external  electric field in the $x$-$z$ plane (the plane containing the tilt vector),  i.e.,  $\mathbf{j}_{x}^{(1)}=\hat{x}\left(4\pi/15v_F\right)\sigma_{0}c_{b}\left(19|\mathbf{E}|B\text{w}_{x}\right)$ but no transverse current density  perpendicular to that plane, i.e., $\mathbf{j}_{y}^{(1)}=0$.

Note that  in the  case of the longitudinal magneto-conductivity set up i.e. when $\mathbf{E}\parallel\mathbf{B}$,  one usually obtains only the longitudinal component of current density
since the Lorentz force is zero. However, in this case, 
we obtain non-zero transverse conductivity in the presence
of tilt and Berry curvature. This unusual transverse signal should be readily measurable in this measurement set-up as one does not need to take into account  the background Lorentz force induced
Hall current. The tilt is a material-specific
parameter whose direction and magnitude  can, in principle, be determined
from the band structure. If these tilt parameters are known, then our model
can provide a prediction of the Berry curvature induced
current. Numerically, from the plot in Fig. \ref{Fig:j1_hall_signal}(d), we predict a planar Hall (PH) conductivity of $14.79$ $\text{oh}\text{m}^{-1}\text{m}^{-1}$
for a tilt value $\text{w}_x=0.1$ along the $x$-direction for Type-I WSM, and  $14.21$ $\text{oh}\text{m}^{-1}\text{m}^{-1}$
for a tilt  $\text{w}_x=1.6$ for Type-II WSM (plot is in the Supplemental). 

Similarly, for the $\mathbf{E}\perp\mathbf{B}$ set-up, we derive the analytical expression for current in the small tilt limit to be,
\begin{align}
\mathbf{j}^{(1)} & =\frac{4\pi}{15}\sigma_{0}c_{b}\left[19\left(\mathbf{E}\cdot\mathbf{w}\right)\mathbf{B}-\left(\mathbf{B}\cdot\mathbf{w}\right)\mathbf{E}\right].\label{eq:firstorderchiral2}
\end{align}

From above, we see  that the current density has no component along the conventional Hall direction (i.e. parallel to $\mathbf{E}\times\mathbf{B}$).  However, there is a non-zero  planar Hall (PH) signal. 
In general,  the PH effect is defined as the   transverse current in the plane containing the electric and magnetic fields  (i.e. along $\mathbf{E}\times(\mathbf{E}\times\mathbf{B})$ direction) when these fields are not parallel to each other.
In our particular scenario, the PH current is parallel to $\mathbf{B}$ and is given by  $\mathbf{j}_{PH}^{(1)}=\left(4\pi/15v_F\right)\sigma_{0}c_{b}19\left(\mathbf{E}\cdot\mathbf{w}\right)\mathbf{B}$. 

\begin{figure}
\begin{centering}
\subfloat[]{\includegraphics[width=0.5\linewidth]{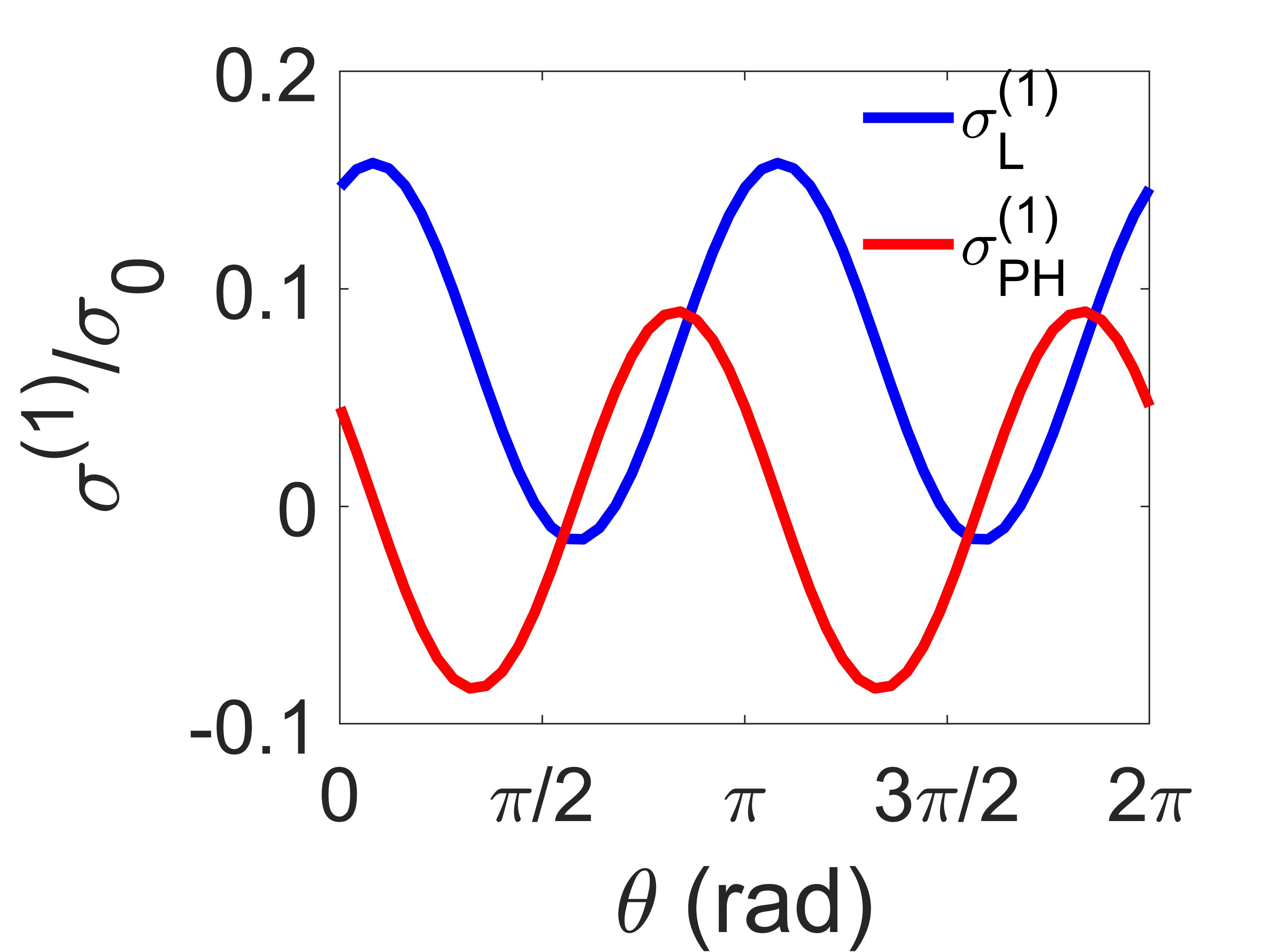}}
\subfloat[]{\includegraphics[width=0.5\linewidth]{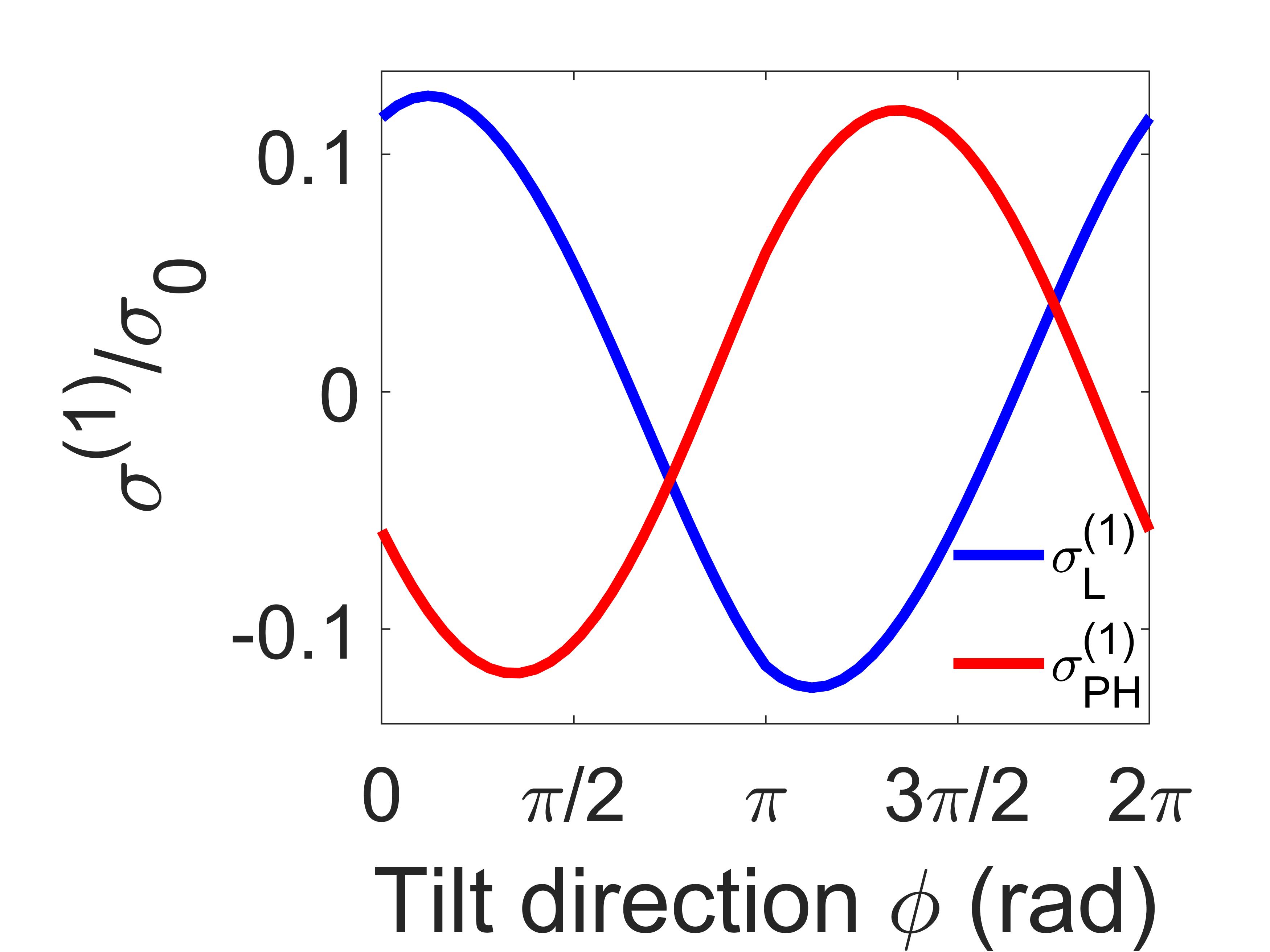}}
\par\end{centering}
\captionsetup{format=plain,indention=0pt,justification=justified}
\caption{\label{Fig:j1_hall_signal-1}(Color online) 
 Plot of longitudinal (blue) and PH (red) components of $\mathbf{j}^{(1)}$ with respect to
 (a)  $\theta$,  the angle between the electric and magnetic fields, and
 (b)  with respect to the tilt direction $\phi$.}
\end{figure}

\textit{Arbitrary tilt strength}:  The analytical current density expressions derived in the previous section holds only in the small tilt limit. For arbitrarily large tilt strength,  we calculate the conductivity numerically based on equations
(\ref{eq:current_zeroth_B}), (\ref{eq:current_first_B}), (\ref{eq:current_second_B}).
In our numerical calculations, we assume  the characteristic collision
time to be $\tau=10^{-13}$ s \cite{Zhang2016,Huang2015} and Fermi
energy $\epsilon_{F}=0.1$ eV. We set the magnetic field to be 3 T, electric
field to be $10^{8}$ V/m and the temperature $T=30$ K.
With these parameter values we have $\sigma_{0}=52.84$ $ \text{Ohm}^{-1}\text{m}^{-1}$.
 In Fig.
\ref{Fig:j1_hall_signal} (a), (b) and (c), we plot the longitudinal components of the zeroth, first and second-order terms of
the current density as a function of the dispersion tilt in the $x$-direction, i.e., 
$\text{w}_{x}$ ($\mathrm{w}_{z}=0$) when the electric field direction is chosen arbitrarily to be $\mathbf{E}\parallel (\hat{x}+\hat{z})$ (i.e. $\mathbf{E}$ lies in the plane containing $\mathbf{w}$ and $\mathbf{B}$). At small tilt, the zeroth $j^{(0)}$
and second-order term $j^{(2)}$ does not change appreciably with the tilt,
confirming our analytical prediction that  the two terms are independent of the tilt at the small tilt limit (Eqs. (\ref{eq:chiraltilt_zeroth})
and (\ref{eq:chiraltilt_second})). However, at sufficiently large
tilt ($|\mathbf{w}|/v_F>0.2$), the response of those terms to the tilt
becomes non-linear. By contrast, the first-order term $j^{(1)}$ varies linearly
with the tilt parameter $\text{w}_{x}$ even at large tilt, a response that is in line with the analytical result of Eq. \eqref{eq:chiraltilt_first}.

Note that all the various orders of  current density show an  increasing trend with the tilt parameter.
A possible reason for this increase of current density is due to the fact that the magnitude of the
group velocity $\mathbf{v_{k}}$ and  the area of the FS both increase
with the tilt strength. 
Although the addition of a tilt term in the Hamiltonian does not change the topological properties of the WSMs,  it does change the energy dispersion and as a consequence the shape of the FS. 
For a Type-I WSM, the shape of FS is ellipsoid where the axis of the ellipsoid is determined by the tilt direction while for Type-II, the shape becomes hyperboloid. 
As the total current density is calculated by integrating over the FS, the shape and size  of the FS have an influence on the current density. 
The effect of the FS geometry is even more prominent in the case of Type-II WSM (this is discussed at the end of this section based on results given in the Supplemental).

Next, we evaluate numerically the transverse current in the longitudinal magneto-conductivity setup, i.e.,  $\mathbf{E}\parallel\mathbf{B}\parallel \hat{z}$. 
In Fig. \ref{Fig:j1_hall_signal}(d), we plot the perpendicular components of
$\mathbf{j}^{(1)}$ corresponding to this set up, both in the  in-plane  ($x$-$z$ plane) and out-of-plane directions, as a function of  the tilt parameter $\text{w}_{x}$. 
In the inset,  the numerical results are compared with the analytical prediction of  Eq. \eqref{eq:firstorderchiral3}, showing good agreement at the small tilt values. 
It can be seen that   the in-plane transverse component of current density  increases linearly with $\text{w}_x$ while the out-of-plane component is zero, both trends of  which  match that of  the analytical form  in Eq.(\ref{eq:firstorderchiral3}).

The presence of the in-plane transverse current in Fig. \ref{Fig:j1_hall_signal}(d) even in the longitudinal set up where the $\mathbf{E}$ and $\mathbf{B}$ fields are parallel to each other can be understood in terms of the group velocity $\mathbf{v}_{\mathbf{k}}$ in Eq. \eqref{eq:current_first_B}.
In the absence of tilt $\mathbf{v}_{\mathbf{k}}\parallel \mathbf{k}\parallel\mathbf{\Omega}_{\mathbf{k}}$,  and, thus the second and third terms are odd in $\mathbf{k}$  as a consequence of which  the integration over FS is zero for those terms, resulting in the current being parallel to $\mathbf{B}$. 
However at finite tilt, $|\mathbf{v_{k}}|= v_F\hat{\mathrm{k}}+\mathbf{w}$, and as a result a component of the current density along the tilt direction appears.
 In our case, we considered $\mathbf{w}=\text{w}_x\hat{x}$ and thus we obtain a transverse current in the $x$-direction, which is also  in accordance with our calculation in Eq. \eqref{eq:firstorderchiral3}.


%

Next, we consider the  longitudinal and PH current densities under varying the angle $\theta$ between the electric and magnetic fields, under varying tilt direction  ( denoted by $\phi$). 
We express the  tilt vector as  $\mathbf{w}=|\mathbf{w}|(\cos\phi~\hat{x}+\sin\phi ~\hat{z})$, where $\phi$ is the angle between the tilt vector and the $x$-axis. 
The electric field is chosen to lie on the $x$-$z$ plane, i.e., $\mathbf{E}=|\mathbf{E}|(\sin\theta~\hat{x}+\cos\theta~\hat{z})$.
Substituting the expressions for  $\mathbf{w}$ and $\mathbf{E}$ into  Eq. \eqref{eq:chiraltilt_first},  the angular dependence of the longitudinal  and PH conductivities at the small tilt is given as,
\begin{equation}
\sigma^{(1)}_{L}=\frac{4 \pi    B |\mathbf{w}| }{15 v_F}\left(19 \sin (2 \theta +\phi )+18 \sin \phi \right),
\end{equation} 
and
\begin{equation}
\sigma^{(1)}_{PH}=\frac{4 \pi  B |\mathbf{w}|}{15 v_F} 19\cos (2 \theta +\phi ).
\end{equation}
Clearly, the  tilt introduces  an additional anisotropy  in both the conductivity terms. 
In Fig. \ref{Fig:j1_hall_signal-1}(a), we plot both the longitudinal and PH conductivities with the variation of the field angle $\theta$  for a fix tilt direction of $\phi=\pi/3$   and magnitude of $|\mathbf{w}|=0.1$. 

In Fig.  \ref{Fig:j1_hall_signal-1}(b) we plot  the variation of both the longitudinal and PH conductivities with the variation of the tilt direction $\phi$, for a fixed  angle between electric and magnetic fields ($\theta=\pi/3$).
 Note that because of the sinusoidal dependence on $\phi$, both current densities change sign, when the tilt direction is reversed, i.e., from $\phi$ to $(\phi+\pi)$. 
 This behavior also matches with the analytical calculation in Eq. \eqref{eq:firstorderchiral3}. Based on this,  let us consider the PH current for different tilt configuration of the Dirac cone for a pair of WPs. 
 Now, if, for a certain value of $\phi$,  the tilt for a pair of WPs is given by $\mathbf{w}_{\pm}=\pm\mathbf{w}$ then the Dirac cones are tilted towards each other. Then for  $(\phi+\pi)$, the tilt vector for the pair  becomes $\mathbf{w}_{\pm}=\mp\mathbf{w}$, i.e., the cones are tilted outwards in k-space (see Fig. \ref{direction1}(c) and (d)) and the direction of the PH current is reversed. 
 This suggests that one can control the PH current  via tilt engineering. For example, if the dispersion tilt can be controlled by  some external parameters such as strain, then one can achieve  strain-induced switching of the PH current or voltage due to the reversal of the relative tilt between the dispersion cones at the two WPs.



\textit{Case (b):} This case corresponds to the scenario where the tilt is uniform, i.e., when $\mathbf{w}_{\chi}=\mathbf{w}$ at all WPs. For this case, the current densities are given by, $\mathbf{j}^{(0)}=(8\pi/3)\sigma_{0}\mathbf{E},$
$\mathbf{j}^{(1)}=0,$ $\mathbf{j}^{(2)}=2\pi\sigma_{0}c_{b}^{2}\left(\frac{1}{3}\left(\mathbf{E}\cdot\mathbf{B}\right)\mathbf{B}+\frac{1}{15}\mathbf{E}\right).$
Specifically, in this case, the PH contribution from the two valleys would cancel out each other giving rise to zero net PH signal. The corresponding 
plots for the tilt dependence of the zeroth and second-order current densities are plotted in the Supplemental.

\textit{Type-II:} For the Type-II WSMs ($|\mathbf{w}|>v_F$) the FS is open and
adopts a hyperboloid configuration. Since, by definition, we are in the large tilt regime, no simple analytical expressions of the current densities can be obtained. 
Instead, we perform numerical calculations of the conductance in Type-II WSM from the integrations in Eqs. (\ref{eq:current_zeroth_B}), (\ref{eq:current_first_B}) and (\ref{eq:current_second_B})  for different values of tilt parameter $\text{w}_{x}$. 
These numerical results  are plotted in the Supplemental.
Interestingly, we find similar trends in the current density compared  to that of Type-I WSM, e.g., the current densities tend to increase  with the tilt strength. 
More importantly, the tilt-induced  PH signal is still present in the case of Type-II WSM.  

\section{Conclusion }
Using a semiclassical approach, we analyze the magnetotransport of a WSM with Dirac cones that are tilted in the momentum space. We calculate the current density up to the second-order in the $B$-field strength and derive the analytical expression in the small tilt limit. The analytical predictions were verified by numerical calculations for arbitrary tilt values. In addition, we evaluate the planar Hall (PH) current when the magnetic field, electric field, and the tilt vector lies in the same plane. Both the longitudinal and PH conductivity can be controlled by tuning the direction and magnitude of the tilt parameter. A tilt-induced transverse current exists even when $\mathbf{E}\parallel\mathbf{B}$,  where the conventional Hall current vanishes. 
This work is supported by the following grants: MOE Tier I (NUS Grant No. R-263-000-D66-114), MOE Tier II MOE2018-T2-2-117 (NUS Grant No. R-398-000-092-112) and NRF-CRP12-2013-01 (NUS Grant No. R-263-000-B30-281).

 \bibliographystyle{unsrt}
\bibliography{draft_WSM_tilted_bibtex}

\end{document}


\preprint{APS/123-QED}

\title{Supplemental material for "Magnetotransport of Weyl semimetals with tilted Dirac cones"}
\author{Anirban Kundu}
\affiliation{Department of Electrical and Computer Engineering, National University of Singapore, 4 Engineering Drive 3, Singapore 117576}
\author{Hyunsoo Yang}
\affiliation{Department of Electrical and Computer Engineering, National University of Singapore, 4 Engineering Drive 3, Singapore 117576}
\author{M. B. A. Jalil}
\affiliation{Department of Electrical and Computer Engineering, National University of Singapore, 4 Engineering Drive 3, Singapore 117576}
\date{\today}


\pacs{}
\maketitle

\section{Type-I analytical calculations}

The total current density is given by, to the linear order in E,
\begin{align}
\mathbf{j} & =\frac{e}{4\pi^{3}\hbar}\int\text{d\ensuremath{\epsilon}}\frac{\partial f^{0}}{\partial\epsilon}\int\text{d}S\frac{\tau(\mathbf{k})}{(1+\frac{e}{\hbar}\mathbf{\Omega}_{\mathbf{k}}\cdot\mathbf{B})}\left[(e\mathbf{E}\cdot\frac{\mathbf{v_{k}}}{|\mathbf{v_{k}}|})\mathbf{v_{k}}+\frac{e^{2}\mathbf{B}}{\hbar}(\mathbf{E}\cdot\frac{\mathbf{v_{k}}}{|\mathbf{v_{k}}|})(\mathbf{\Omega}_{\mathbf{k}}\cdot\mathbf{v_{k}})\right.\nonumber \\
 & \left.+\frac{e^{2}}{\hbar}(\mathbf{E}\cdot\mathbf{B})\left\{ (\mathbf{\Omega}_{\mathbf{k}}\cdot\frac{\mathbf{v_{k}}}{|\mathbf{v_{k}}|})\mathbf{v_{k}}+\frac{e\mathbf{B}}{\hbar^{2}}(\mathbf{\Omega}_{\mathbf{k}}\cdot\frac{\mathbf{v_{k}}}{|\mathbf{v_{k}}|})^{2}|\mathbf{v_{k}}|\right\} \right]\label{eq:main}
\end{align}
We expand to the orders in magnetic field B, 
\begin{equation}
\frac{1}{(1+\frac{e}{\hbar}\mathbf{\Omega}_{\mathbf{k}}\cdot\mathbf{B})}\approx1-\frac{e}{\hbar}\mathbf{\Omega}_{\mathbf{k}}\cdot\mathbf{B}+(\frac{e}{\hbar}\mathbf{\Omega}_{\mathbf{k}}\cdot\mathbf{B})^{2}+\mathcal{O}\left(|\mathbf{B}|^{3}\right)
\end{equation}
Separating the terms to the orders in $\mathbf{B}$ we get the
total current density to be,

\begin{align}
\mathbf{j} & =\mathbf{j}^{(0)}+\mathbf{j}^{(1)}+\mathbf{j}^{(2)}
\end{align}
Small-tilt limit:\textit{ }for small tilt, we do the following to
expand each term to the linear order in tilt,

\begin{align}
\mathbf{v_{k}} & =(\mathbf{w}+v_F\hat{\mathrm{k}})=\left(v_F\hat{\mathrm{k}}+\mathbf{w}\right),\text{ }\text{ }\\
\mathbf{v_{k}}/|\mathbf{v_{k}}| & =\left.\hat{\mathrm{k}}+\left(\frac{\mathbf{w}}{v_F}-(\hat{\mathrm{k}}\cdot\frac{\mathbf{w}}{v_F})\hat{\mathrm{k}}\right)\right.,
\end{align}
such that for any vector $\mathbf{A}$,
\begin{align}
\mathbf{A}\cdot\frac{\mathbf{v_{k}}}{|\mathbf{v_{k}}|}\mathbf{v_{k}} & =v_F(\mathbf{A}\cdot\hat{\mathrm{k}})\hat{\mathrm{k}}+(\mathbf{A}\cdot\hat{\mathrm{k}})\mathbf{w}+\mathbf{A}\cdot(\mathbf{w}-(\hat{\mathrm{k}}\cdot\mathbf{w})\hat{\mathrm{k}})\hat{\mathrm{k}}.
\end{align}
Using the above, the zeroth order term,
\begin{align}
\mathbf{j}^{(0)} & =\frac{e\tau}{4\pi^{3}\hbar}\int\text{d\ensuremath{\epsilon}}\frac{\partial f^{0}}{\partial\epsilon}\int\text{d}S\text{ }\left[...\right]\\
\text{ }\left[...\right] & =v_Fe\left[(\mathbf{E}\cdot\hat{\mathrm{k}})\hat{\mathrm{k}}\right]+e\left[(\mathbf{E}\cdot\hat{\mathrm{k}})\mathbf{w}+\mathbf{E}\cdot(\mathbf{w}-(\hat{\mathrm{k}}\cdot\mathbf{w})\hat{\mathrm{k}})\hat{\mathrm{k}}\right].\nonumber 
\end{align}
The first order term,

\begin{align}
\mathbf{j}^{(1)} & =\frac{e\tau}{4\pi^{3}\hbar}\int\text{d\ensuremath{\epsilon}}\frac{\partial f^{0}}{\partial\epsilon}\int\text{d}S\text{ }\left[...\right].\\
\left[...\right] & =\frac{e^{2}}{\hbar}\left[v_F\mathbf{B}(\mathbf{E}\cdot\hat{\mathrm{k}})(\mathbf{\Omega}_{\mathbf{k}}\cdot\hat{\mathrm{k}})+\left(\mathbf{E}\cdot\mathbf{B}\right)v_F(\mathbf{\Omega}_{\mathbf{k}}\cdot\hat{\mathrm{k}})\hat{\mathrm{k}}-\left(\mathbf{\Omega}_{\mathbf{k}}\cdot\mathbf{B}\right)\left\{ v_F\left[(\mathbf{E}\cdot\hat{\mathrm{k}})\hat{\mathrm{k}}\right]\right\} \right]^{(0^{th}\text{in tilt})}\nonumber \\
\\
 & +\frac{e^{2}}{\hbar}\left[\mathbf{B}\left\{ (\mathbf{E}\cdot\hat{\mathrm{k}})(\mathbf{\Omega}_{\mathbf{k}}\cdot\mathbf{w})+\left(\mathbf{E}\cdot(\mathbf{w}-(\hat{\mathrm{k}}\cdot\mathbf{w})\hat{\mathrm{k}})\right)(\mathbf{\Omega}_{\mathbf{k}}\cdot\hat{\mathrm{k}})\right\} \right.\nonumber \\
 & \left.+\left(\mathbf{E}\cdot\mathbf{B}\right)\left\{ \left[\mathbf{w}\mathbf{\Omega}_{\mathbf{k}}\cdot\hat{\mathrm{k}}+\mathbf{\Omega}_{\mathbf{k}}\cdot(\mathbf{w}-(\hat{\mathrm{k}}\cdot\mathbf{w})\hat{\mathrm{k}})\hat{\mathrm{k}}\right]\right\} -\left(\mathbf{\Omega}_{\mathbf{k}}\cdot\mathbf{B}\right)\left\{ \left[(\mathbf{E}\cdot\hat{\mathrm{k}})\mathbf{w}+\mathbf{E}\cdot(\mathbf{w}-(\hat{\mathrm{k}}\cdot\mathbf{w})\hat{\mathrm{k}})\hat{\mathrm{k}}\right]\right\} \right]^{(1^{st}\text{in tilt})}\nonumber 
\end{align}
As one can immediately verify that $0^{\text{th}}$ in tilt terms
vanishes after the constant surface (spherical in shape) integration
and only 1st order in tilt terms survives. The second order term,
\begin{align}
\mathbf{j}^{(2)} & =\frac{e\tau}{4\pi^{3}\hbar}\int\text{d\ensuremath{\epsilon}}\frac{\partial f^{0}}{\partial\epsilon}\int\text{d}S\text{ }\left[...\right]\\
\left[...\right] & =e^{2}\left[\left(\mathbf{E}\cdot\mathbf{B}\right)\mathbf{B}v_Fe(\mathbf{\Omega}_{\mathbf{k}}\cdot\hat{\mathrm{k}})^{2}-e\mathbf{B}(\mathbf{\Omega}_{\mathbf{k}}\cdot\mathbf{B})v_F(\mathbf{E}\cdot\hat{\mathrm{k}})(\mathbf{\Omega}_{\mathbf{k}}\cdot\hat{\mathrm{k}})\right.\nonumber \\
 & \left.\text{- }\left(\mathbf{E}\cdot\mathbf{B}\right)e\left\{ (\mathbf{\Omega}_{\mathbf{k}}\cdot\mathbf{B})v_F(\mathbf{\Omega}_{\mathbf{k}}\cdot\hat{\mathrm{k}})\hat{\mathrm{k}}\right\} +(\mathbf{\Omega}_{\mathbf{k}}\cdot\mathbf{B})^{2}ev_F(\mathbf{E}\cdot\hat{\mathrm{k}})\hat{\mathrm{k}}\right]^{(0^{th}\text{in tilt})}\nonumber \\
\nonumber \\
 & +e^{2}\left[e\left(\mathbf{E}\cdot\mathbf{B}\right)\mathbf{B}\left\{ (\mathbf{\Omega}_{\mathbf{k}}\cdot\hat{\mathrm{k}})^{2}(\hat{\mathrm{k}}\cdot\mathbf{w})+2(\mathbf{\Omega}_{\mathbf{k}}\cdot\hat{\mathrm{k}})\mathbf{\Omega}_{\mathbf{k}}\cdot(\mathbf{w}-(\hat{\mathrm{k}}\cdot\mathbf{w})\hat{\mathrm{k}})\right\} \right.\nonumber \\
 & -\mathbf{B}e(\mathbf{\Omega}_{\mathbf{k}}\cdot\mathbf{B})\left\{ (\mathbf{E}\cdot\hat{\mathrm{k}})(\mathbf{\Omega}_{\mathbf{k}}\cdot\mathbf{w})+\left(\mathbf{E}\cdot(\mathbf{w}-(\hat{\mathrm{k}}\cdot\mathbf{w})\hat{\mathrm{k}})\right)(\mathbf{\Omega}_{\mathbf{k}}\cdot\hat{\mathrm{k}})\right\} \nonumber \\
 & \text{-}\left(\mathbf{E}\cdot\mathbf{B}\right)e\left\{ (\mathbf{\Omega}_{\mathbf{k}}\cdot\mathbf{B})\left[(\mathbf{\Omega}_{\mathbf{k}}\cdot\hat{\mathrm{k}})\mathbf{w}+\mathbf{\Omega}_{\mathbf{k}}\cdot(\mathbf{w}-(\hat{\mathrm{k}}\cdot\mathbf{w})\hat{\mathrm{k}})\hat{\mathrm{k}}\right]\right\} \nonumber \\
 & \left.+(\mathbf{\Omega}_{\mathbf{k}}\cdot\mathbf{B})^{2}\left\{ e\left(\left[(\mathbf{E}\cdot\hat{\mathrm{k}})\mathbf{w}+\mathbf{E}\cdot(\mathbf{w}-(\hat{\mathrm{k}}\cdot\mathbf{w})\hat{\mathrm{k}})\hat{\mathrm{k}}\right]\right)\right\} \right]^{(1^{st}\text{in tilt})}
\end{align}

All the terms to the first order in tilt above vanishes after integration
over the Fermi surfaces.Now the above result is for single valley.
To get the total effect of tilt we sum over two valleys with opposite
chirality ($\chi=\pm1$ ) where the Berry curvature is given by $\mathbf{\Omega}_{\mathbf{k}}=\chi\left(\mathrm{k}_{F}^{2}/2\mathrm{k}^{3}\right)\mathbf{k}$.
While doing so we consider two cases: a) the tilt changes sign between
the valleys with opposite chirality i.e. $\mathbf{w}_{\chi}=\chi\mathbf{w}$,
(b) the sign of the tilt remains unchanged in the two valleys $\mathbf{w}_{\chi}=\mathbf{w}$.
Clearly, in all cases only the terms with even order in $\chi$ survives.
Considering spherical Fermi surface i.e. $\text{d}S_F=k_F^{2}\sin^{2}\theta\text{d}\theta\text{d}\phi $
and $\partial f^{0}/\partial\epsilon=\theta(\epsilon-\epsilon_{F})$
we obtain following results,

for case (a): 
\begin{align}
\mathbf{j}^{(0)} & =\frac{e^{2}v_F\tau k_{F}^{2}}{4\pi^{3}\hbar}\left(2\times\frac{4\pi}{3}\mathbf{E}\right),
\end{align}
\begin{align*}
\mathbf{j}^{(1)} & =2\times\frac{e^{3}\tau}{4\pi^{3}\hbar}\int\text{d\ensuremath{\epsilon}}\frac{\partial f^{0}}{\partial\epsilon}\text{ }\left[...\right],\\
\left[...\right] & =\int\text{d}S\\
 & \left[\mathbf{B}\left\{ (\mathbf{E}\cdot\hat{\mathrm{k}})(\mathbf{\Omega}_{\mathbf{k}}\cdot\mathbf{w})+\left(\mathbf{E}\cdot(\mathbf{w}-(\hat{\mathrm{k}}\cdot\mathbf{w})\hat{\mathrm{k}})\right)(\mathbf{\Omega}_{\mathbf{k}}\cdot\hat{\mathrm{k}})\right\} \right.\\
 & \left.+\left(\mathbf{E}\cdot\mathbf{B}\right)\left\{ \mathbf{w}\mathbf{\Omega}_{\mathbf{k}}\cdot\hat{\mathrm{k}}+\mathbf{\Omega}_{\mathbf{k}}\cdot(\mathbf{w}-(\hat{\mathrm{k}}\cdot\mathbf{w})\hat{\mathrm{k}})\hat{\mathrm{k}}\right\} -\left(\mathbf{\Omega}_{\mathbf{k}}\cdot\mathbf{B}\right)\left((\mathbf{E}\cdot\hat{\mathrm{k}})\mathbf{w}+\mathbf{E}\cdot(\mathbf{w}-(\hat{\mathrm{k}}\cdot\mathbf{w})\hat{\mathrm{k}})\hat{\mathrm{k}}\right)\right]\\
 & =2\pi\mathbf{B}\left(\mathbf{E}\cdot\mathbf{w}\right)+2\pi\left(\mathbf{E}\cdot\mathbf{B}\right)\mathbf{w}+\frac{2\pi}{15}B\left[\hat{x}\left(3\mathbf{E}\cdot\hat{z}\mathrm{w}_{x}+\mathbf{E}\times\mathbf{w}\cdot\hat{y}\right)-\hat{y}\left(\mathbf{E}\times\mathbf{w}\cdot\hat{x}\right)+\hat{z}\left(4\mathbf{E}\cdot\mathbf{w}+3\mathbf{E}\cdot\hat{z}\mathrm{w}_{z}\right)\right]\\
 & =\frac{2\pi}{15}B\left[\hat{x}\left(18\mathbf{E}\cdot\hat{z}\mathrm{w}_{x}+\mathbf{E}\times\mathbf{w}\cdot\hat{y}\right)-\hat{y}\left(\mathbf{E}\times\mathbf{w}\cdot\hat{x}\right)+\hat{z}\left(19\mathbf{E}\cdot\mathbf{w}+18\mathbf{E}\cdot\hat{z}\mathrm{w}_{z}\right)\right]\\
 & =\frac{2\pi}{15}\left[18\left(\mathbf{E}\cdot\mathbf{B}\right)\mathbf{w}+19\left(\mathbf{E}\cdot\mathbf{w}\right)\mathbf{B}+\hat{x}\left(\mathbf{E}\times\mathbf{w}\cdot\hat{y}\right)B-\hat{y}\left(\mathbf{E}\times\mathbf{w}\cdot\hat{x}\right)B\right]\\
\mathbf{j}^{(1)} & =\sigma_{0}c_{b}\frac{4\pi}{15}\left[18\left(\mathbf{E}\cdot\mathbf{B}\right)\mathbf{w}+19\left(\mathbf{E}\cdot\mathbf{w}\right)\mathbf{B}+\hat{x}\left(\mathbf{E}\times\mathbf{w}\cdot\hat{y}\right)B-\hat{y}\left(\mathbf{E}\times\mathbf{w}\cdot\hat{x}\right)B\right],\\
\end{align*}

\begin{align}
\mathbf{j}^{(2)} & =2\times\frac{e\tau}{4\pi^{3}\hbar}\int\text{d\ensuremath{\epsilon}}\frac{\partial f^{0}}{\partial\epsilon}\int\text{d}S\text{ }\left[...\right],\\
\left[...\right] & =\frac{e^{2}}{\hbar^{2}}\left[\left(\mathbf{E}\cdot\mathbf{B}\right)\mathbf{B}v_Fe(\mathbf{\Omega}_{\mathbf{k}}\cdot\hat{\mathrm{k}})^{2}-e\mathbf{B}(\mathbf{\Omega}_{\mathbf{k}}\cdot\mathbf{B})v_F(\mathbf{E}\cdot\hat{\mathrm{k}})(\mathbf{\Omega}_{\mathbf{k}}\cdot\hat{\mathrm{k}})\right.\nonumber \\
 & \left.\text{- }\left(\mathbf{E}\cdot\mathbf{B}\right)e\left\{ (\mathbf{\Omega}_{\mathbf{k}}\cdot\mathbf{B})v_F(\mathbf{\Omega}_{\mathbf{k}}\cdot\hat{\mathrm{k}})\hat{\mathrm{k}}\right\} +(\mathbf{\Omega}_{\mathbf{k}}\cdot\mathbf{B})^{2}ev_F(\mathbf{E}\cdot\hat{\mathrm{k}})\hat{\mathrm{k}}\right]^{(0^{th}\text{in tilt})}\nonumber \\
 & =e^{3}v_F\frac{\pi}{\hbar^{2}k^{2}}\left[\frac{1}{3}\left(\mathbf{E}\cdot\mathbf{B}\right)\mathbf{B}+\frac{1}{15}\mathbf{E}\right]\\
\mathbf{j}^{(2)} & =2\pi\sigma_{0}c_{b}^{2}\left[\frac{1}{3}\left(\mathbf{E}\cdot\mathbf{B}\right)\mathbf{B}+\frac{1}{15}\mathbf{E}\right],
\end{align}

similarly, for case (b): 
\begin{align}
\mathbf{j}^{(0)} & =2\times\frac{4\pi}{3}\mathbf{E},
\end{align}
\begin{align}
\mathbf{j}^{(1)} & =0,
\end{align}
\begin{align}
\mathbf{j}^{(2)} & =2\pi\sigma_{0}c_{b}^{2}\left[\frac{1}{3}\left(\mathbf{E}\cdot\mathbf{B}\right)\mathbf{B}+\frac{1}{15}\mathbf{E}\right].
\end{align}

\section{type-I, non-chiral tilt:}

\begin{figure}
\begin{centering}
\subfloat[]{\includegraphics[width=0.4\linewidth]{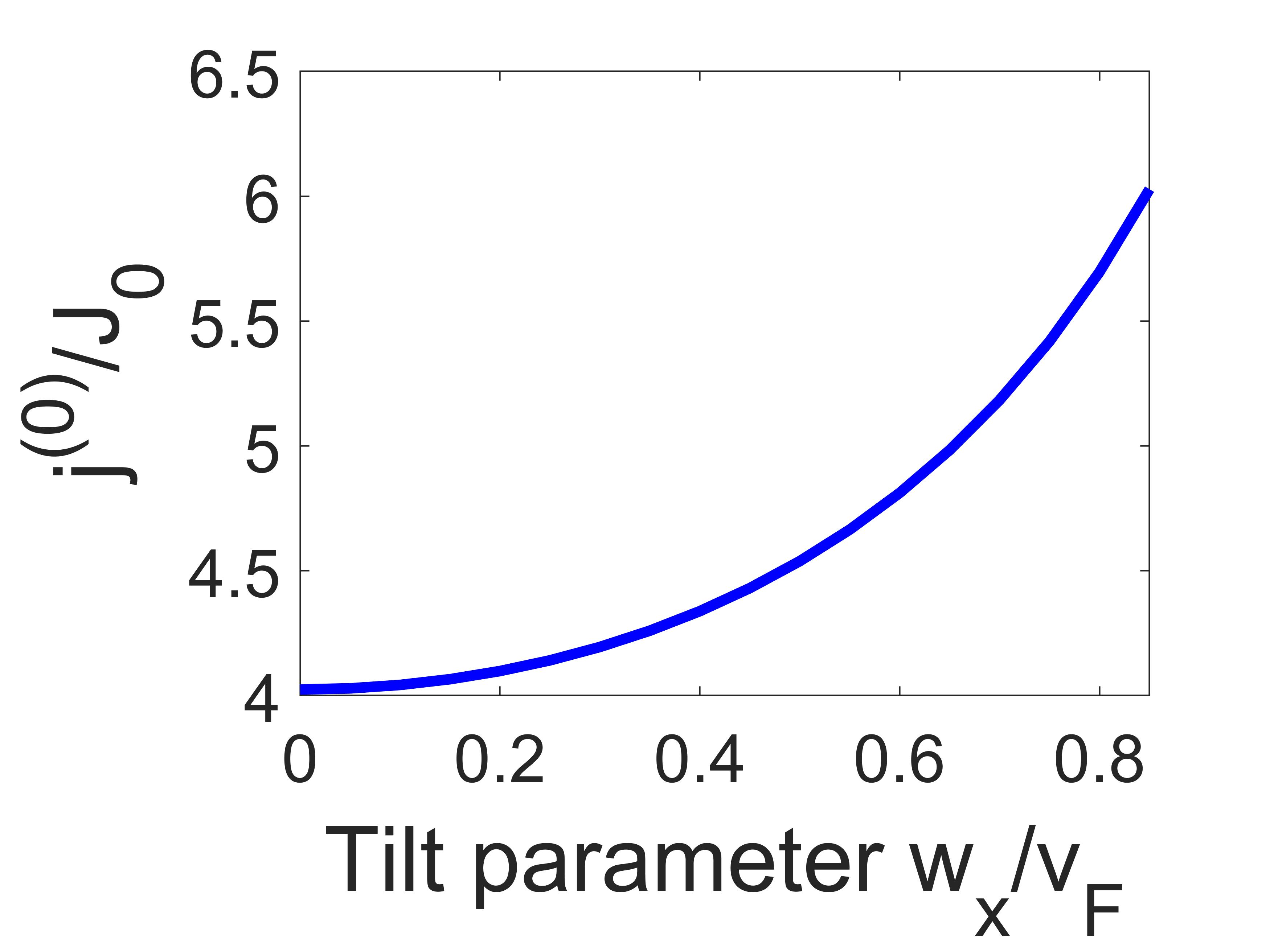}}\subfloat[]{\includegraphics[width=0.4\linewidth]{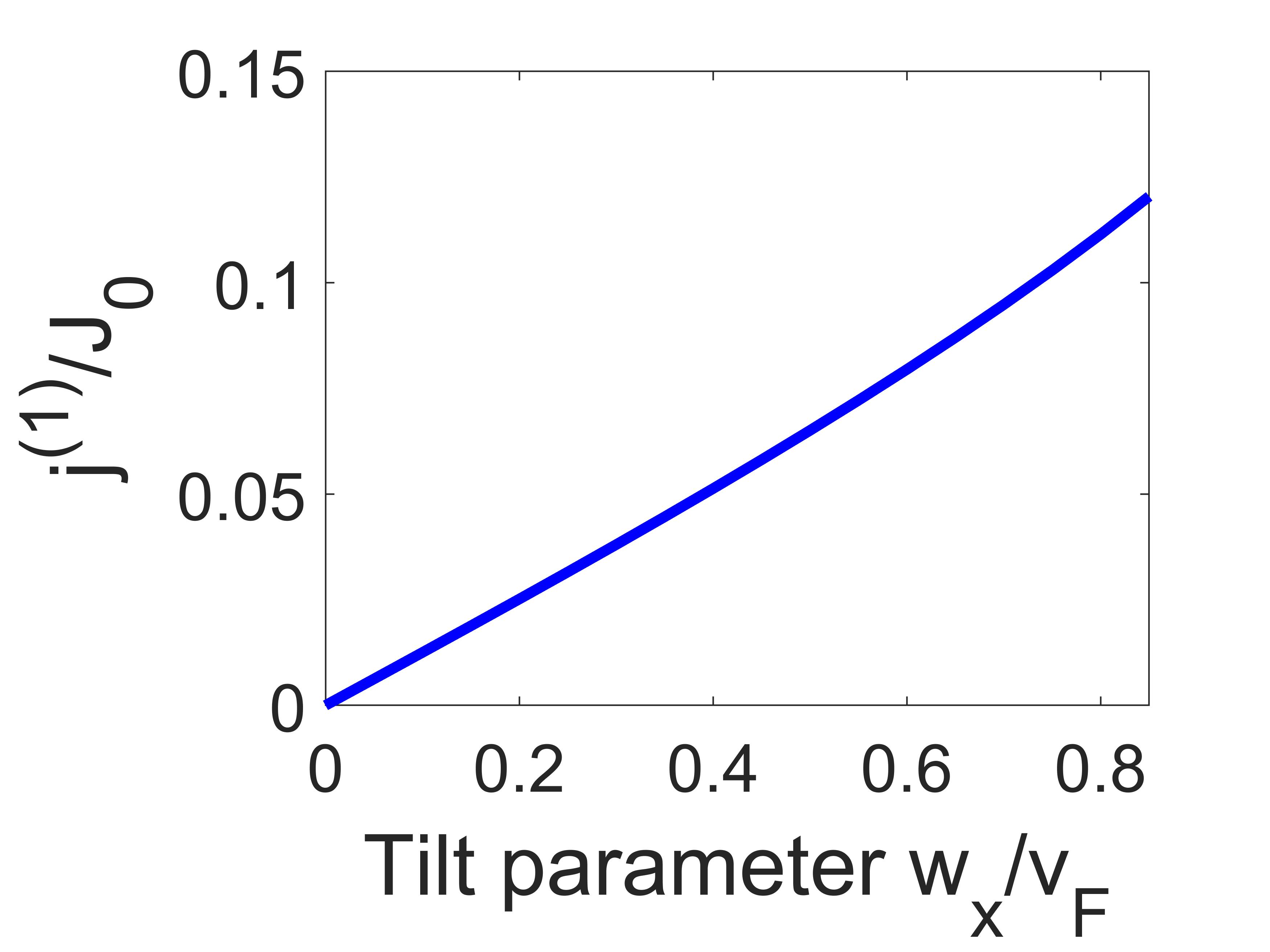}}
\par\end{centering}
\begin{centering}
\vfill{}
\subfloat[]{\includegraphics[width=0.4\linewidth]{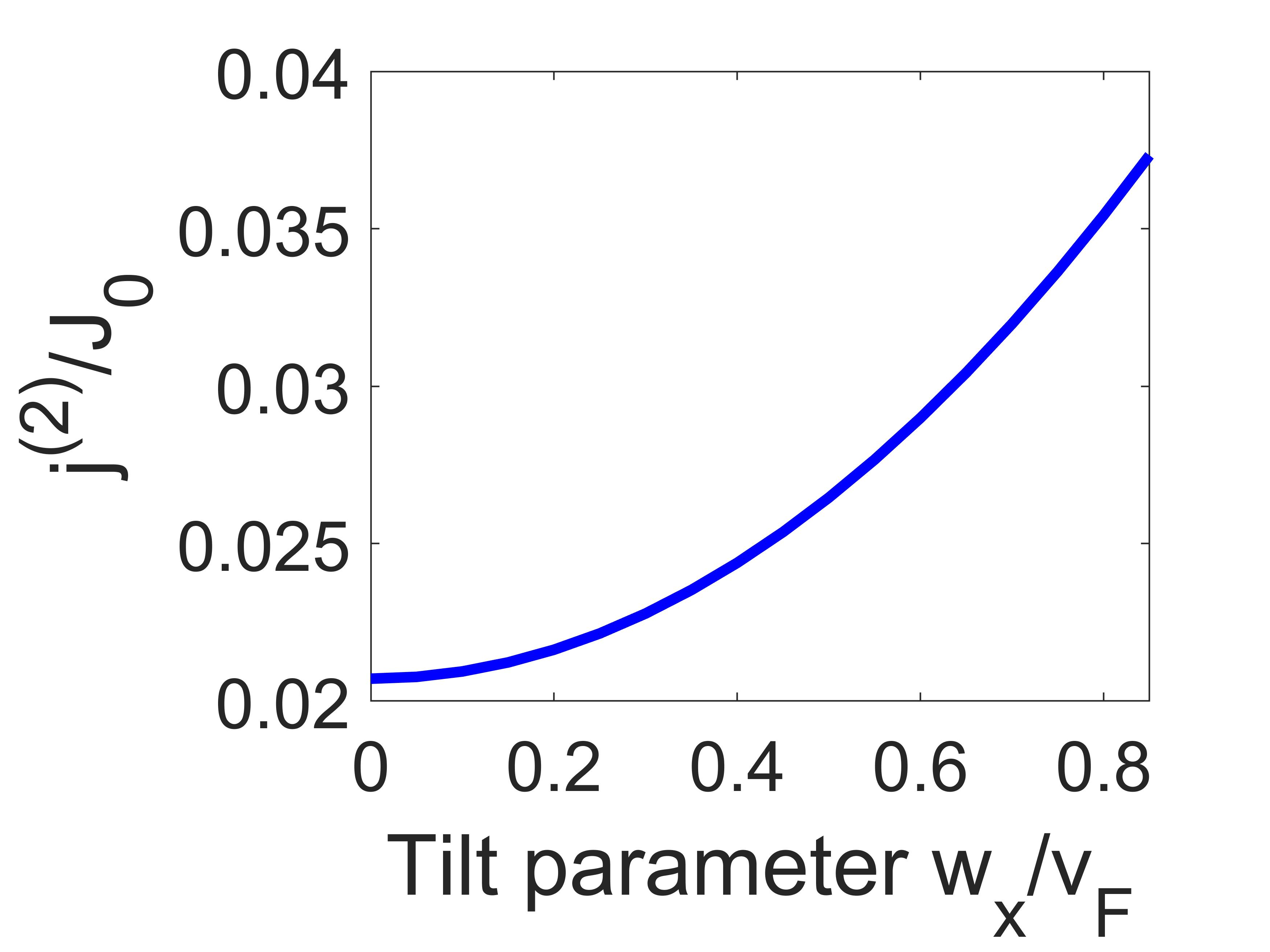}}
\subfloat[]{\includegraphics[width=0.4\linewidth]{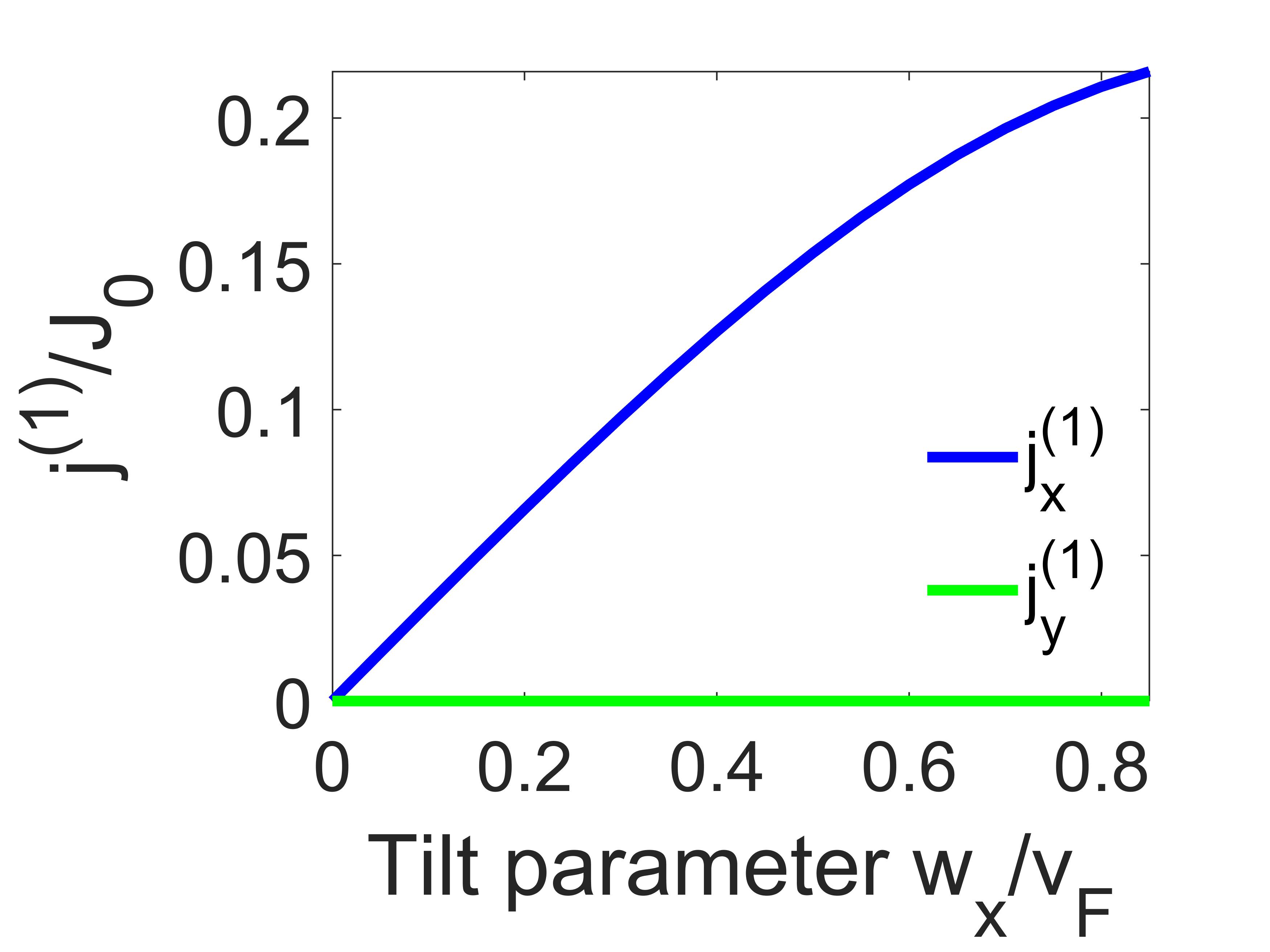}}
\par\end{centering}
\caption{\label{Fig:S_j1_hall_signal} Plot of longitudinal components of current
density for, (a) $\mathbf{j}^{(0)}$, (b) $\mathbf{j}^{(1)}$,
(c) $\mathbf{j}^{(2)}$ with respect to the $x$-component of
tilt $\mathrm{w}_{x}$ ($\mathrm{w}_{z}=0$) when electric field $\mathbf{E}$ along
$(\hat{x}+\hat{z})$ direction. (d) Plot of $x$ and $y$
components of $\mathbf{j}^{(1)}$ with respect to $x$-component
of tilt $\text{w}_{x}$ when $\mathbf{E}||\mathbf{B}||\hat{z}$. 
 }
\end{figure}

\begin{figure}
\begin{centering}
\subfloat[]{\includegraphics[width=0.4\linewidth]{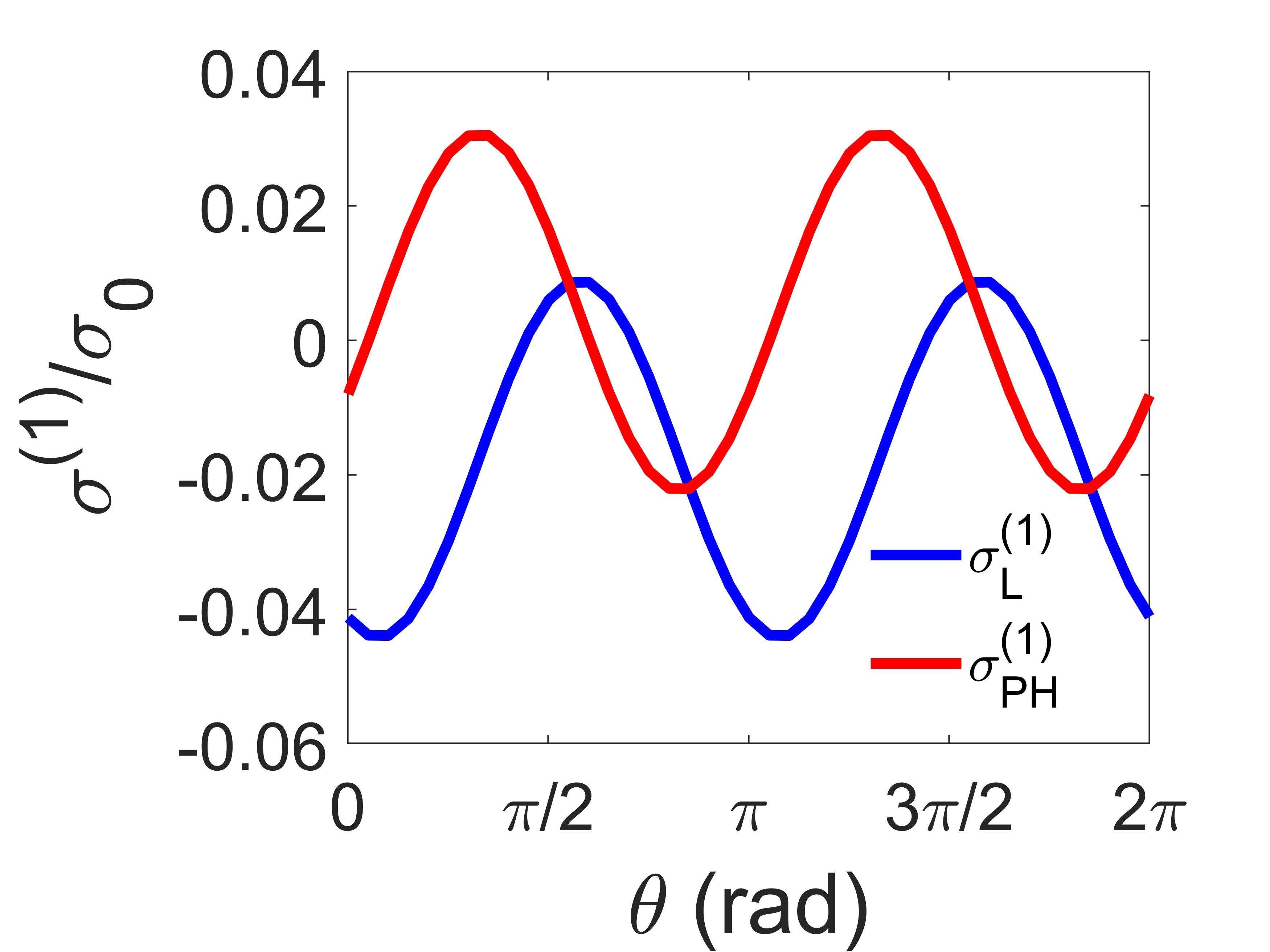}}
\subfloat[]{\includegraphics[width=0.4\linewidth]{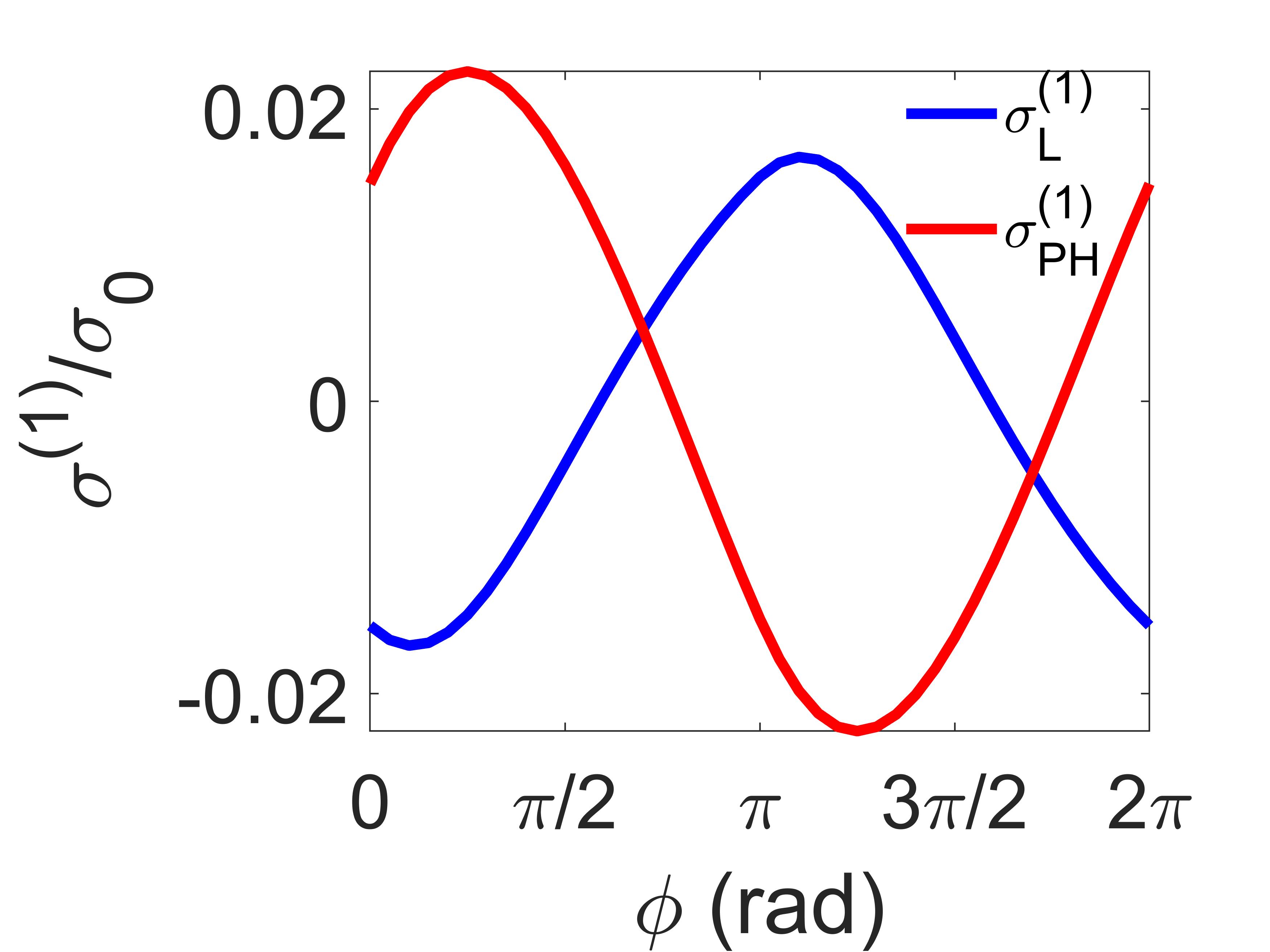}}
\par\end{centering}
\caption{\label{Fig:S_j1_hall_signal-1a} 
Plot of longitudinal (blue) and PH (red) components of $\mathbf{j}^{(1)}$ with respect to
 (a)  $\theta$,  the angle between the electric and magnetic fields, and
 (b)  with respect to the tilt direction $\phi$.
}
\end{figure}

Here we investigate the case (b) i.e. when the sign of the tilt does not alter between valleys with opposite chiralities. We integrate Eqs.
(6)-(8) (as given in the manuscript) numerically for different values of $\mathrm{w}_{x}$ ($\mathrm{w}_{z}=0$). 
The longitudinal components of $\mathbf{j}^{(0)}$, $\mathbf{j}^{(1)}$, $\mathbf{j}^{(2)}$ when the electric field direction is specified in the literature are given in Fig. \ref{Fig:S_j1_hall_signal}(a)-(c),  and, the  transverse currents when the electric field is parallel to magnetic field is plotted in Fig. \ref{Fig:S_j1_hall_signal}(d). 
The variation of longitudinal and Planar Hall (PH) components for the current density $\mathbf{j}^{(1)}$ with $\theta$  i.e.  the angle between $\mathbf{E}$ and $\mathbf{B}$  are
given in Fig.  \ref{Fig:S_j1_hall_signal-1a}(a) (when we fix $\phi=\pi/3$ and $|\mathbf{w}|=0.1$) and the variation with the angle $\phi$ i.e. the angle between tilt direction and $x$-axis are given in \ref{Fig:S_j1_hall_signal-1a}(b) (when we fix $\theta=\pi/3$ and $|\mathbf{w}|=0.1$).

\section{type-II, chiral tilt}

\begin{figure}
\begin{centering}
\subfloat[]{\includegraphics[width=0.40\linewidth]{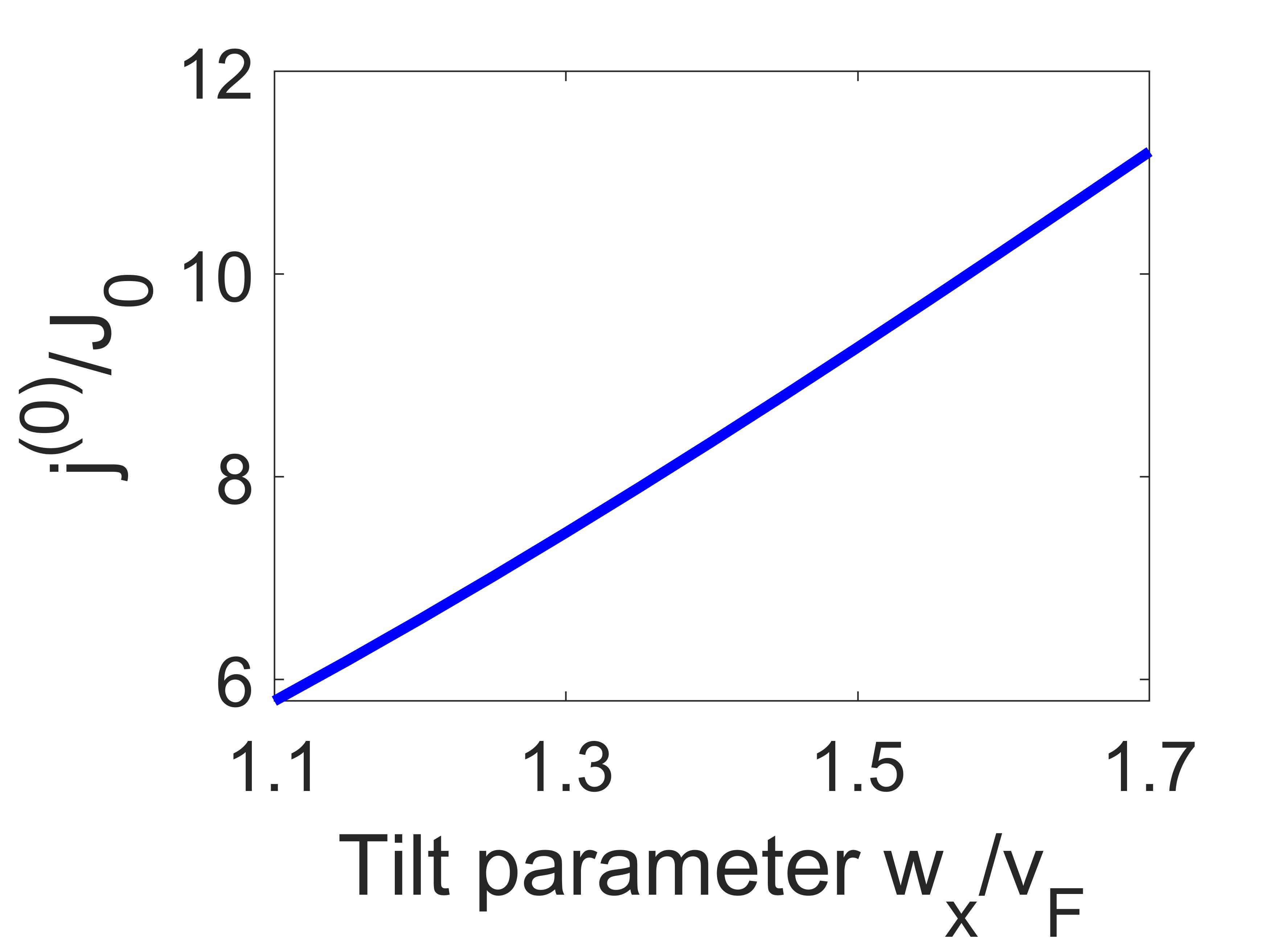}}
\subfloat[]{\includegraphics[width=0.40\linewidth]{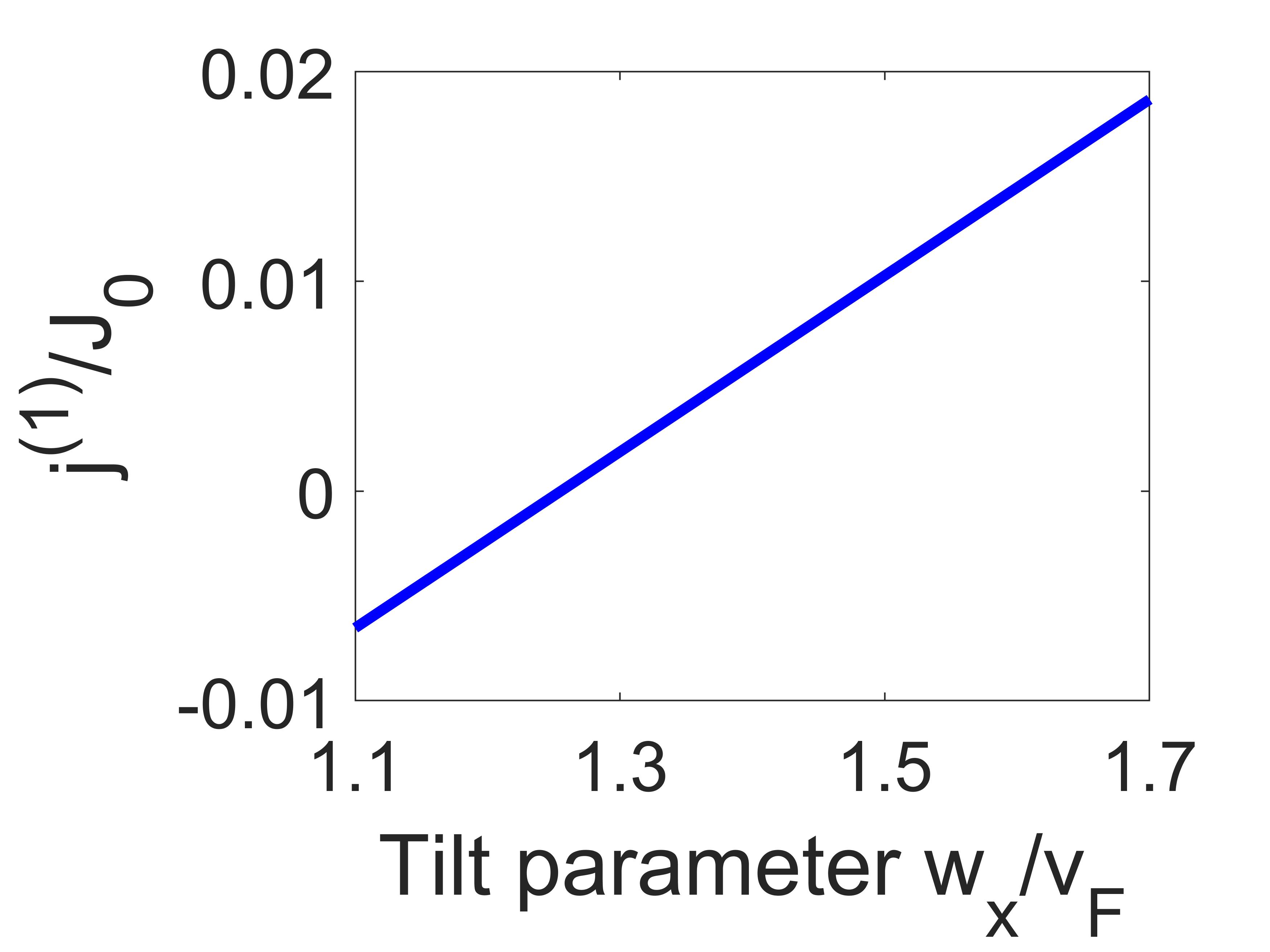}}
\vfill{}
\par
\subfloat[]{\includegraphics[width=0.4\linewidth]{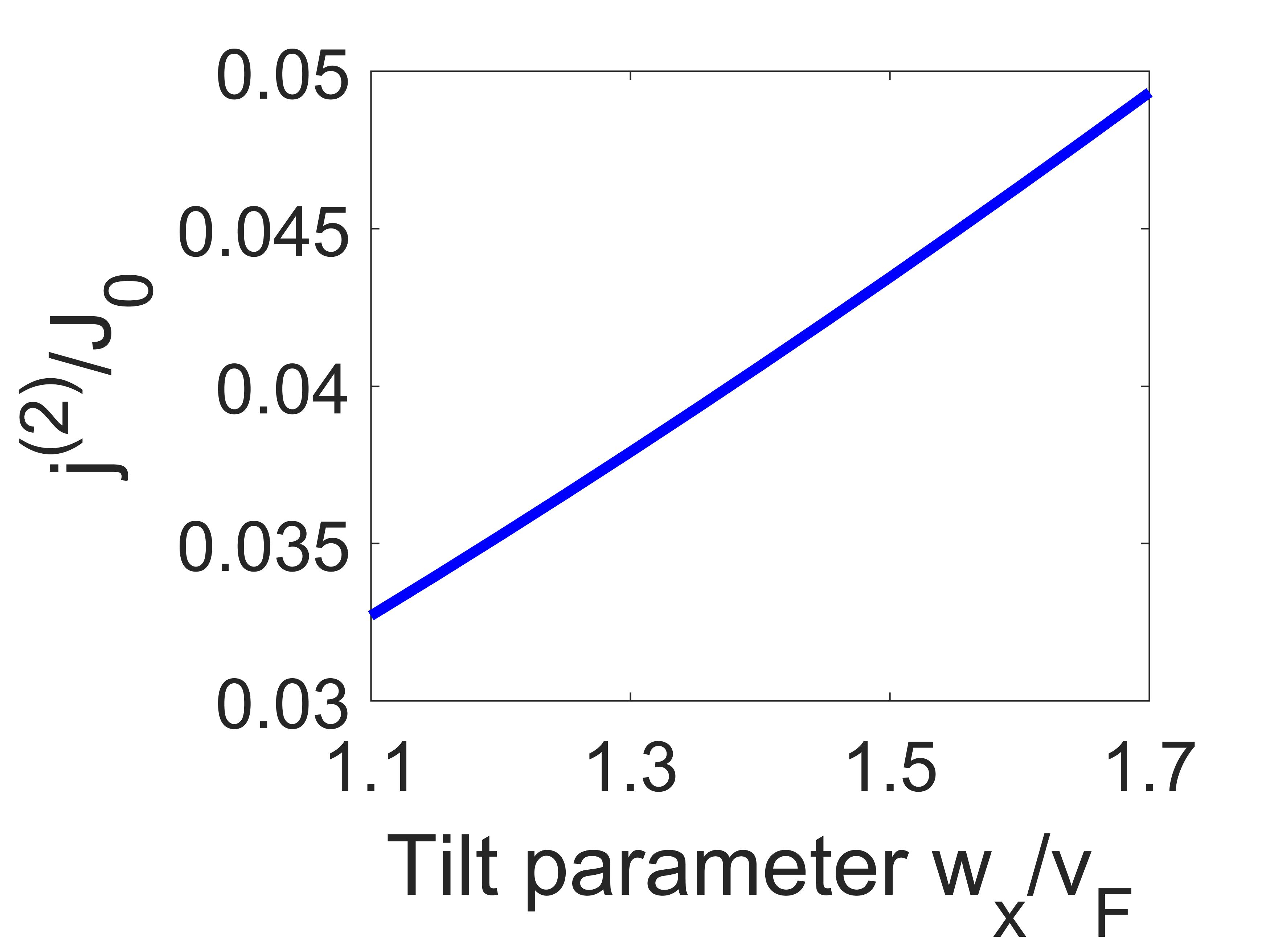}}
\subfloat[]{\includegraphics[width=0.4\linewidth]{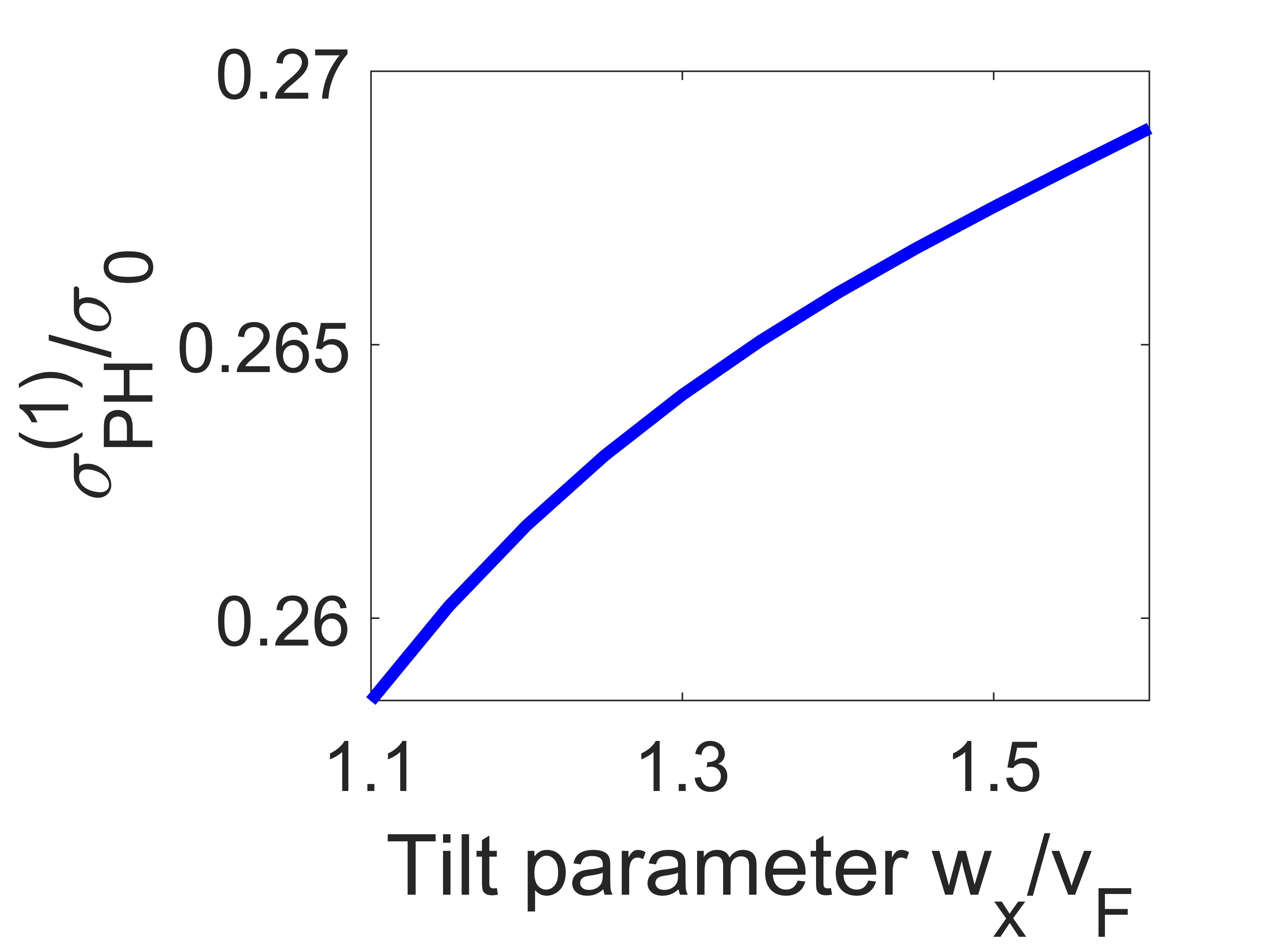}}
\par\end{centering}
\caption{\label{Fig: typeII_j1_hall_signal} For Type-II WSM: Plot of longitudinal
component of current density, (a) for $\mathbf{j}^{(0)}$, (b)
$\mathbf{j}^{(1)}$, (c) $\mathbf{j}^{(2)}$ with respect
to the $x$-component of tilt $\mathrm{w}_{x}$ ($\mathrm{w}_{z}=0$) when electric
field $\mathbf{E}$ along $(\hat{x}+\hat{z})$ direction. (d):
Plot of Planar Hall component (along $\hat{x}$-direction) of current
density $\mathbf{j}^{(1)}$ with respect to $\mathrm{w}_{x}$ when $\mathbf{E}||\mathbf{B}||\hat{z}$. }
\end{figure}

\begin{figure}
\begin{centering}
\subfloat[]{\includegraphics[width=0.40\linewidth]{j1_angle_E_and_B.jpeg}}
\subfloat[]{\includegraphics[width=0.40\linewidth]{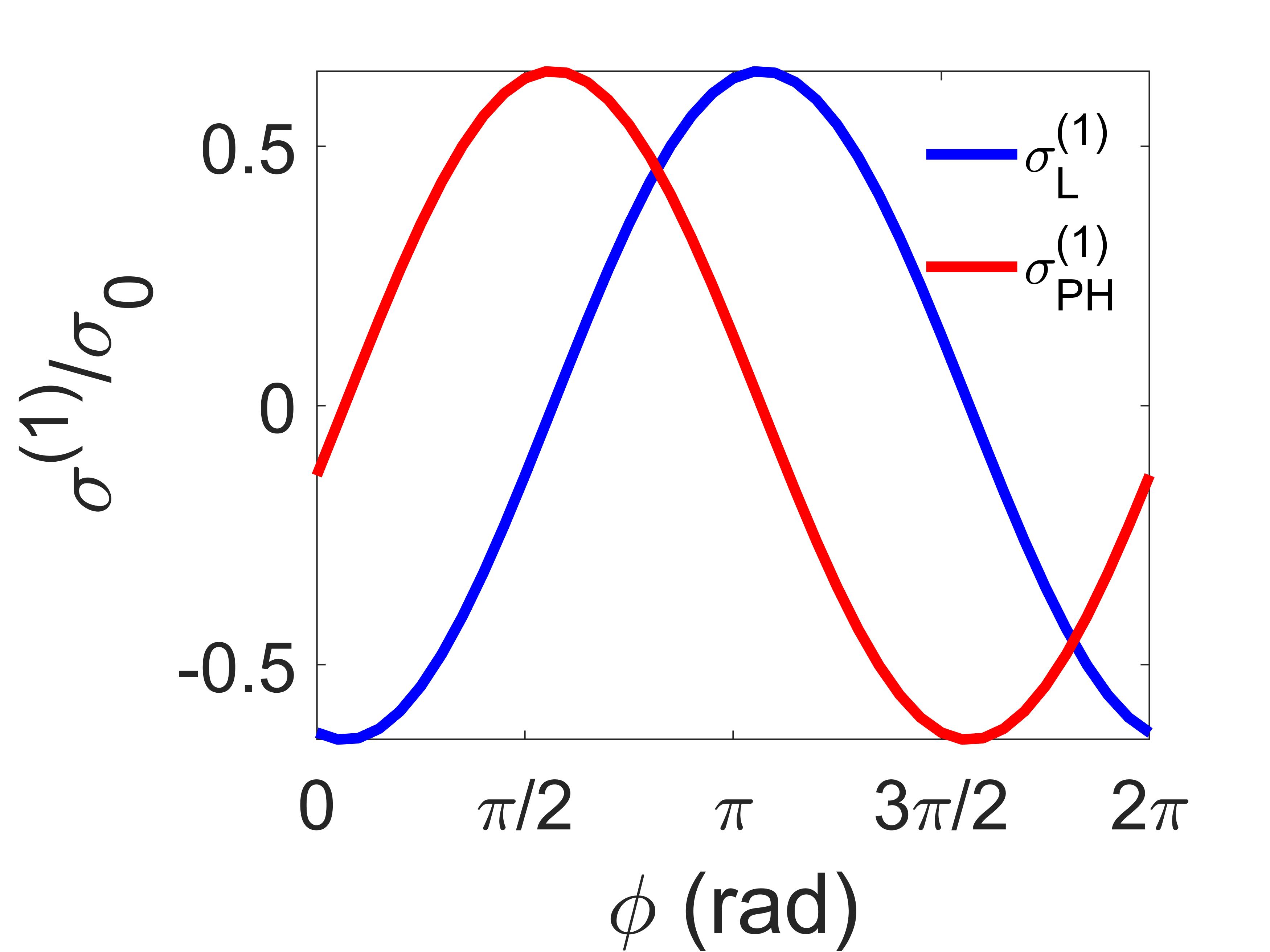}}
\par\end{centering}
\caption{\label{Fig:S_j1_hall_signal-1-1} 
Plot of longitudinal (blue) and PH (red) components of $\mathbf{j}^{(1)}$ with respect to
 (a)  $\theta$,  the angle between the electric and magnetic fields, and
 (b)  with respect to the tilt direction $\phi$.
}
\end{figure}

Similar to the type-I WSM (as described in the manuscript) the variation with the tilt parameter $\mathrm{w}_{x}$ for type-II WSM are
given in the Fig. \ref{Fig: typeII_j1_hall_signal}(a)-(d). The variation of longitudinal and PH components for the current density $\mathbf{j}^{(1)}$ with $\theta$  (i.e.  the angle between $\mathbf{E}$ and $\mathbf{B}$)  are
given in Fig.  \ref{Fig:S_j1_hall_signal-1-1}(a) (when we fix $\phi=\pi/3$ and $|\mathbf{w}|=1.3$) and  with the angle $\phi$ i.e. the angle between tilt direction and $x$-axis are given in \ref{Fig:S_j1_hall_signal-1-1}(b) (when we fix $\theta=\pi/3$ and $|\mathbf{w}|=1.3$).

